\documentclass[%reprint,
superscriptaddress,%groupedaddress,%unsortedaddress,%runinaddress,%frontmatterverbose,
showpacs,%preprintnumbers,floatfix,
nofootinbib,%nobibnotes,%bibnotes,
amsmath,amssymb,%aps,
pra,%prb,
twocolumn
]{revtex4-2}

\usepackage{graphicx}% Include figure files
\usepackage[all,arc,knot,matrix,color]{xy}
\usepackage{amsthm}
\usepackage{braket}
\usepackage{natbib}
\usepackage{hyperref}
\hypersetup{colorlinks=true, linkcolor=blue, citecolor=blue, urlcolor=blue}
\usepackage{tikz}
\usepackage[oldstyle]{libertine}
\usepackage{libertinust1math}

\newcommand{\avg}[1]{\langle #1\rangle}

\begin{document}

\title{Jensen-Mercer two-sided bounds for  entropy production rate}
\author{Xiao-Kan Guo}
 \email{kankuohsiao@whu.edu.cn}
\affiliation{School of Mathematics \& Physics, Yancheng Institute of Technology, Yancheng 224051, Jiangsu, China}
\author{Zhiqiang Huang}\email{zqhuang@hubu.edu.cn}
\affiliation{School of Physics, Hubei University, Wuhan 430062, Hubei, China}
\date{\today}

\begin{abstract}
Bounds on entropy production convert limited observations into constraints on hidden
dissipation.  The Jensen bound expresses a lower bound on the total and subsystem entropy
production rates of continuous stochastic systems in terms of mean coordinate rates alone.
We show that the same convexity structure yields a complementary upper bound, inspired by Mercer's reflection inequality.
 For the quadratic entropy-production functional of overdamped Langevin dynamics, Mercer's reflection inequality is identical to the sharp Bhatia-Davis variance bound and coincides with the
convex dual of the smallest affine majorant of the functional over the range of the local
current.  The two bounds yield a two-sided sandwich controlled solely by the mean drift and
the extrema of the current, organized into a hierarchy of optimal bounds in which each further
observation tightens the ceiling. The
thermodynamic uncertainty relation supplies an independent fluctuation floor.   We
also treat position-dependent diffusion, underdamped dynamics and the tilted periodic
potential, for which the weak-barrier series is obtained in closed form and the strong-barrier
regime is solved asymptotically. We check the bounds numerically on the Brownian motor.
\end{abstract}
\maketitle

%%%%%%%%%%%%%%%%%%%%%%%%%%%%%%%%%%%%
\section{Introduction}
Bounding and estimating entropy production is a central goal of nonequilibrium
thermodynamics \cite{Seifert12,Seifert19,Parrondo15}.  The second law supplies a lower bound
on the average entropy production rate.  The sharper, observable-based bounds are provided by
the thermodynamic uncertainty relation (TUR) \cite{BH15,HG15,Gingrich16,Pietzonka16,
Proesmans17}, by the  geometric and speed-limit bounds \cite{Shiraishi18,Okuyama18}, and by the
Jensen bound introduced for multicomponent stochastic systems in Refs.~\cite{LS22,LS24}. 
The
Jensen bound lower-bounds the total and subsystem entropy production rates in terms of bare
diffusion coefficients and mean coordinate rates, and has been used to constrain
collective motor-driven transport and to infer free-energy transduction in bipartite
molecular machines.
 For one-dimensional driven transport a counterpart that bounds the current fluctuations from
above has been conjectured \cite{GuZhang25}, which motivates the upper bounds on the entropy
production rate studied here.

The Jensen bound follows from Jensen's inequality \cite{Jensen06} applied to the quadratic
entropy-production functional of the local probability current.   The purpose of the
present work is to exploit the same convexity at its natural second stage, the reflection
inequality of Mercer \cite{Mercer03}, and to determine what a ceiling of this kind can and
cannot deliver.  For the quadratic functional the reflection inequality and the chord
(Bhatia--Davis) inequality \cite{BD00}, which sharpens the classical variance bound of
Popoviciu \cite{Pop35}, are algebraically the same statement.   Averaging the chord inequality  pointwise yields the sharp ceiling, attained on a two-point
current and equal to the dual value of the smallest affine majorant of the functional over
the range of the local current.  On the other hand, averaging the reflection inequality and then estimating the
reflected second moment  yields a weaker Mercer-type ceiling.  
This is the bound produced by the affine Jensen--Mercer family, and we show
that no interior member of that family improves on the ceiling.  
We therefore obtain  a sandwich bounding  the entropy
production rate between a fluctuation-free floor and an extremal-fluctuation ceiling, both
expressible through the mean drift and the range of the local current.

To find the sharp bound in general cases, we map the ceiling to the
solution of a moment problem in linear programming. Then, different ranges of the local current will lead to different sharp upper bounds which can be solved by the linear optimization.
In particular, the stationarity constrains
the local current to a periodic solution of a Riccati equation, so the distributional ceiling is
attained only by a constant current, while a bound on the drive lowers it explicitly.
The tilted periodic (washboard) potential \cite{Reimann01,Reimann02,Haenggi09}, which is the
standard model of a Brownian motor and whose strong-barrier limit is the Kramers escape
problem \cite{Kramers40} and whose equilibrium diffusivity is the Lifson--Jackson value
\cite{Lifson62}, quantifies the deficit. It is small at weak barrier and diverges at strong
barrier, where the surviving floor is the fluctuation one.  The TUR floor deviates from the
exact rate only at $O(g^{2})$, whereas the Jensen floor decays exponentially in the
strong-barrier regime, where the TUR floor survives at $\tanh(\pi a)/(\pi a)$ of the rate.

The optimization method has been previously applied to obtain a strict lower bound on the entropy production \cite{Skinner21}, in which the dispersion of an statistical estimator is turned into a confidence statement.
The optimization of the upper bounds of the present work is of a different type.  
They are deterministic,  instead of statistical, worst-case
statements, requiring only that the local current lie in a known range.
  Throughout, the range is treated as data, and when it is
itself estimated with an error $\delta$, the ceiling widens by $\delta$.

We begin in Section~\ref{sec:setup} by reviewing the Jensen lower bound for an overdamped stochastic system.
Section~\ref{sec:mercer} derives the sharp upper bound, separates it from the weaker bound
obtained by averaging Mercer's reflection inequality, and studies its linear-programming dual.
Section~\ref{sec:obs} develops the optimal bounds, the TUR floor, the dynamical attainability of
the ceiling in one dimension.  Section~\ref{5} extends the bounds to position-dependent
diffusion, underdamped dynamics and a tilted periodic potential, where the fixed-tilt realizability
of the data is settled exactly.
Section~\ref{sec:model} checks the analytic results numerically on the Brownian motor, and
Sec.~\ref{sec:summary} concludes.

%%%%%%%%%%%%%%%%%%%%%%%%%%%%%%%%%%%%
\section{Setup and the Jensen lower bound}\label{sec:setup}
We consider an overdamped stochastic system with $N$ degrees of freedom
$x=\{x_i\}_{i=1}^N$, each coupled to a heat bath at temperature $T_i$, with
diagonal mobility, friction, and diffusion tensors.  In terms of the
probability density $p(x,t)$ the Fokker--Planck equation reads 
\begin{equation}\partial_t p=-\nabla\!\cdot\!J\end{equation}
where  $J=\mu f p-D\nabla p$ is the  probability flux with mobility $\mu$, external force $f$, and diffusion coefficient $D$\cite{Risken,Kampen}.  For this system, the entropy
production rate (scaled by the Boltzmann constant
$k_B$) is \cite{LS24,Sekimoto}
\begin{equation}
\dot\sigma=\sum_{i=1}^N D_i^{-1}\,\big\langle y_i^{2}\big\rangle,
\label{eq:EP}
\end{equation}
where $y_i(x,t)\equiv J_i(x,t)/p(x,t)$ is the local probability-current
velocity and
\begin{equation}
\avg{y_i}=\int\frac{J_i}{p}\,p\,dx=\int J_i\,dx=v_i.
\end{equation}
Throughout the paper $\langle\cdot\rangle$ denotes an average over the steady-state density
$p(x)$. Spatial (or angular) averages over the ring are always written out explicitly, as in
$(2\pi)^{-1}\int_0^{2\pi}(\cdot)\,dx$.

 Because $D^{-1}$ is positive semidefinite, the function $Q(y)=y\!\cdot\!D^{-1}y$ is convex, so Jensen's inequality  gives
\begin{equation}
\dot\sigma_i=D_i^{-1}\avg{y_i^{2}}\ \ge\ D_i^{-1}\avg{y_i}^{2}=D_i^{-1}v_i^{2}.
\label{eq:Ajensen}
\end{equation}
This gives the Jensen
bound \cite{LS22,LS24}
\begin{equation}
\dot\sigma\ \ge\ \sum_{i=1}^N D_i^{-1}\,v_i^{2}
\label{eq:jensen}
\end{equation}
This is a lower bound on the  entropy production rates in terms of bare
diffusion coefficients $D$ and mean coordinate rates $v_i$.

 For general diffusion tensors, $Q(y)=y\!\cdot\!D^{-1}y$ is
convex for positive-semidefinite $D^{-1}$, and  $\dot\sigma$ reads
$\dot\sigma=\langle y\!\cdot\!D^{-1}y\rangle$, of which \eqref{eq:EP} is the diagonal case.
 Jensen's inequality gives
$\dot\sigma\ge\avg{y}\!\cdot\!D^{-1}\avg{y}$.

%%%%%%%%%%%%%%%%%%%%%%%%%%%%%%%%%%%%
\section{Mercer's inequality and the sharp upper bound}\label{sec:mercer}
We assume that the local current velocity is bounded in an interval, i.e.
$y_i(x,t)\in[\alpha_i,\beta_i]$ almost everywhere.  For any $y\in[\alpha,\beta]$, the following
identity holds
\begin{equation}
\alpha^2+\beta^2-y^2-(\alpha+\beta-y)^2=2\,(y-\alpha)(\beta-y)\ \ge\ 0.
\label{eq:mercerBD}
\end{equation}
It is Mercer's reflection inequality \cite{Mercer03} for $y^{2}=f(y)$
and, equivalently, the pointwise Bhatia--Davis inequality \cite{BD00},
\begin{equation}y^{2}\le(\alpha+\beta)y-\alpha\beta.\label{7}
\end{equation}  

The two forms \eqref{eq:mercerBD} and \eqref{7} lead to two different bounds, and we keep them apart.  Averaging the
\emph{chord} form \eqref{7} over $p$ with $\avg{y}=v$ yields the sharp upper bound
\begin{equation}
\dot\sigma\ \le\ \sum_{i=1}^N
\frac{(\alpha_i+\beta_i)\,v_i-\alpha_i\beta_i}{D_i}\equiv U_{\mathrm{BD}}.
\label{eq:upper}
\end{equation}
 Equation~(\ref{eq:upper}) is the exact maximum of the average $\langle Q\rangle$
over the distributions supported in the box $\mathcal B=\prod_{i=1}^{N}[\alpha_i,\beta_i]$ with mean $v$.
For $N=1$ the maximizing measure is unique.  For $N>1$ the value is still exact, because $Q$ is
separable and the marginal means fix it, while only the coordinate marginals are determined.  A
non-diagonal $D^{-1}$ couples the components, and then the optimum follows from the linear program
of Appendix~\ref{app:LP}.
 Nevertheless, a continuous current that
takes only the two endpoint values is constant, so equality forces $\alpha_i=\beta_i=v_i$.  The
bound is attained only by the degenerate constant current and is strict for every non-degenerate
steady state.

Averaging instead the \emph{reflection} form,
$\avg{y^{2}}+\avg{(\alpha+\beta-y)^{2}}\le\alpha^{2}+\beta^{2}$, and lower-bounding the
reflected second moment by Jensen's inequality,
$\avg{(\alpha+\beta-y)^{2}}\ge(\alpha+\beta-v)^{2}$, gives the weaker Mercer-type ceiling
\begin{equation}
\begin{split}
\dot\sigma_i\ \le\ U_{\mathrm{Merc}}&\equiv
\sum_i\frac{\alpha_i^{2}+\beta_i^{2}-(\alpha_i+\beta_i-v_i)^{2}}{D_i}\\
&\equiv U_{\mathrm{BD}}+\sum_i\frac{\mathrm{Var}_{\max,i}}{D_i},
\end{split}
\label{eq:mercertype}
\end{equation}
which exceeds the sharp ceiling by exactly the maximal variance
$\mathrm{Var}_{\max,i}\equiv(v_i-\alpha_i)(\beta_i-v_i)$.  The gap
$U_{\mathrm{Merc}}-U_{\mathrm{BD}}$ is the excess of estimating the reflected second moment
instead of using the chord form directly.  It vanishes only at the endpoints of the admissible
mean, $v=\alpha$ or $v=\beta$, and it is exactly the bound produced by the affine
Jensen--Mercer family at its upper endpoint
(Sec.~\ref{sec:affine}).

In variance form the sharp bound \eqref{eq:upper} reads
\begin{equation}0\le\mathrm{Var}\,y_i\le(v_i-\alpha_i)(\beta_i-v_i)=
\mathrm{Var}_{\max},\end{equation}
so that both ceilings are statements about the second moment of the local current at fixed
mean and support, differing by exactly one maximal variance.

 Together these give the
two-sided sandwich for the entropy production rate,
\begin{equation}
\sum_i\frac{v_i^{2}}{D_i}\ \le\ \dot\sigma\ \le\
\sum_i\frac{(\alpha_i+\beta_i)v_i-\alpha_i\beta_i}{D_i}.
\label{eq:sandwich}
\end{equation}
whose ends are the Jensen floor \eqref{eq:jensen} and the sharp ceiling \eqref{eq:upper}.

%%%%%%%%%%%%%%%%%%%%%%%%%%%%%%%%%%%%
\subsection{General diffusion tensor and linear programming}\label{sec:general}
For a general positive-semidefinite $D^{-1}$ the sharp upper bound is the smallest affine
majorant of $Q$ over the box $\mathcal B=\prod_{i=1}^{N}[\alpha_i,\beta_i]$, evaluated at $v$,
\begin{equation}
\dot\sigma\ \le\ L^\ast(v),\qquad
L^\ast=\inf\{\,L\ \text{affine}\ :\ L\ge Q\ \text{on Box}\,\},
\label{eq:LP}
\end{equation}
which by convexity equals the linear-programming dual over the box vertices $c^{(s)}$
[Appendix~\ref{app:LP}],
\begin{equation}
L^\ast(v)=\min_{\mu_0,\lambda}\Big\{\mu_0+\lambda\!\cdot\! v\ :\
\mu_0+\lambda\!\cdot\! c^{(s)}\ge Q(c^{(s)}),
s=1,\dots,2^{N}\Big\}.
\label{eq:dual}
\end{equation}
It is exact for every $N$, but
evaluating it requires the $2^{N}$ vertices of the box, so it is a characterization rather than a
closed form.  
Here, by exact we mean the value of this program. 

Diagonalizing $D^{-1}=U\Lambda U^\top$ and applying Eq.~(\ref{eq:mercerBD}) mode by mode
gives an explicit affine majorant,
\begin{equation}
\dot\sigma\ \le\ M(v)=\sum_k\lambda_k\big[(\alpha^{(k)}+\beta^{(k)})(u_k\!\cdot\!v)
-\alpha^{(k)}\beta^{(k)}\big]\ \ge\ L^\ast(v),
\label{eq:eigen}
\end{equation}
with $\alpha^{(k)}=\sum_i\min(u_{k,i}\alpha_i,u_{k,i}\beta_i)$ and $\beta^{(k)}$ defined
analogously.  The coordinated Jensen--Mercer inequality \cite{Toseef25,Nwaeze26} gives the
looser four-corner bound
\begin{equation}
\dot\sigma\le\sum_{c\in\mathrm{corners}}Q(c)-Q(\alpha+\beta-v).
\end{equation}

For example, if $D^{-1}=\bigl(\begin{smallmatrix}2&1\\1&2\end{smallmatrix}\bigr)$ and the box is
$[0,1]^{2}$, the value function is explicitly
$L^\ast(v)=2[v_1+v_2+\min(v_1,v_2)]$, a concave piecewise-linear function whose two pieces
are glued along the diagonal $v_1=v_2$.  At $v=(\tfrac12,\tfrac12)$ it
takes the value $3$, attained by the affine majorant $3(y_1+y_2)$ and by the two-point
current $\tfrac12\delta_{(0,0)}+\tfrac12\delta_{(1,1)}$.  The same majorant gives $3$ at
$v=(1,0)$, where the envelope is $2$ and the optimal majorant is $2y_1+4y_2$, so a single
supporting hyperplane is not a substitute for the envelope.  At $v=(\tfrac12,\tfrac12)$ the
eigenmode bound, the four-corner bound, and the Jensen floor take the values $7/2$, $17/2$,
and $3/2$, so that it verifies, $\tfrac32\le\dot\sigma\le3\le\tfrac72\le\tfrac{17}{2}$. See Appendix~\ref{app:LP} for details.

%%%%%%%%%%%%%%%%%%%%%%%%%%%%%%%%%%%%
\section{Optimal bounds }\label{sec:obs}
\subsection{The sharp ceiling bounds}
The box assumption of Sec.~\ref{sec:mercer} is one piece of prior information about the
local current velocity, and each further piece yields its own optimal ceiling.  For a
family $\mathcal C$ of admissible distributions with mean $v$ the tightest bound is
\begin{equation}
\dot\sigma\ \le\ U^\ast(v;\mathcal C),\qquad
U^\ast(v;\mathcal C)=\sup_{\mu\in\mathcal C}\int Q\,d\mu ,
\label{eq:hier}
\end{equation}
which is nonincreasing as $\mathcal C$ shrinks.  The pieces of information are of two kinds.  A \emph{distributional} datum constrains the law of $y$.   A \emph{dynamical} datum restricts the drift that produces the law.  It does not shrink
the distributional class $\mathcal C$, although it restricts the subset of $\mathcal C$ that a
steady state can realize.

 In the following, the data are assumed \emph{admissible}:
$\alpha\le v\le\beta$ for any declared range, $\epsilon\ge0$ and $\beta\le R$ for a tail mass,
and $m_2\ge v^{2}$ for a second moment.  If a datum is not admissible the class $\mathcal C$ is
empty and no ceiling exists.

For the one-dimensional quadratic diffusion tensor
$Q=y^{2}/D$, with the extension to $N$ independent modes understood, the natural classes
admit closed forms:
\begin{itemize}
\item[(i)] mean alone, $U^{\ast}=\infty$.
\item[(ii)] mean and range $y\in[\alpha,\beta]$, the Bhatia--Davis value
$U_{\mathrm{BD}}=[(\alpha+\beta)v-\alpha\beta]/D$. This is the case of Sec.~\ref{sec:mercer}. 
\item[(iii)] mean and one-sided range $y\in[0,\beta]$, $U_1=\beta v/D$, which for
$0\le\alpha\le v$ exceeds $U_{\mathrm{BD}}$ by $\alpha(\beta-v)/D\ge0$ and for $\alpha<0$
is the tighter of the two. This class presupposes $y\ge0$, so it applies to a current of
known sign and not to a general one.
\item[(iv)] mean, range and a bounded tail mass: Knowing that the mass above $\beta$ is bounded, and
$y\in[\alpha,R]$ almost surely with $\Pr(y>\beta)\le\epsilon$ and $\alpha\le v\le\beta\le R$,
adds a three-point extremal bound
\begin{equation}
U_\epsilon=U_{\mathrm{BD}}+\frac{1}{D}\min\{\epsilon(R-\alpha)(R-\beta),\
(v-\alpha)(R-\beta)\},
\label{eq:tail}
\end{equation}
which reduces to $U_{\mathrm{BD}}$ as $\epsilon\to0$.  Here $\epsilon$ \emph{bounds} the tail
mass,  while $R$ is the endpoint of the admissible range.  
\item[(v)] mean and a bound on the drift, $|f|\le F_{\max}$ or
$f\in[f_{-},f_{+}]$. This is not a distributional datum.  It leaves $\mathcal C$ unchanged and restricts the
dynamically realizable subset.
\end{itemize}
The ceilings (ii) and (iii) are attained on two points, $\{\alpha,\beta\}$ and
$\{0,\beta\}$. See Appendix \ref{app:obs} for details.

When the mean and the second moment
$\langle y^{2}\rangle=m_2$ are known, the objective is fixed exactly [Appendix \ref{app:obs}]
\begin{equation}U^{\ast}=m_2/D=\dot\sigma,\end{equation} although the underlying distribution is not. The classes (ii) and (iv) are therefore the natural finite-precision
interpolations between the box ceiling and this exact determination.

%%%%%%%%%%%%%%%%%%%%%%%%%%%%%%%%%%%%
\subsection{The thermodynamic uncertainty relation as a fluctuation floor}\label{sec:tur}
The Jensen lower bound uses only the mean current.  When the current fluctuates, its accumulated
variance carries additional information.  For a steady current $J_t$ with mean $j=\langle
J\rangle$ and long-time variance $2D_J t$, the finite-time TUR
\cite{BH15,HG15} reads $\mathrm{Var}(J_T)/\langle J_T\rangle^{2}\ge 2/(\dot\sigma T)$.  In the long-time
limit $\langle J_T\rangle\simeq jT$ and $\mathrm{Var}(J_T)\simeq 2D_JT$, so
\begin{equation}
\dot\sigma\ \ge\ \frac{j^{2}}{D_J},
\label{eq:tur}
\end{equation}
with unit coefficient, the time canceling between the two sides.  For the tilted
periodic potential the steady displacement is $j=v$ and $D_J$ is the effective diffusion
coefficient $D_{\mathrm{eff}}$ of the process \cite{Reimann01,Reimann02}.  Hence
\begin{equation}
\dot\sigma\ \ge\ L_{\mathrm{TUR}}\equiv\frac{v^{2}}{D_{\mathrm{eff}}},\qquad
L_{\mathrm{Jensen}}=\frac{v^{2}}{D},
\label{eq:turfloor}
\end{equation}
and the TUR floor dominates the Jensen floor exactly when $D_{\mathrm{eff}}<D$, that is,
when the tilted periodic potential suppresses the current fluctuations below free
diffusion.  
%%%%%%%%%%%%%%%%%%%%%%%%%%%%%%%%%%%%
\subsection{Affine Jensen--Mercer family}\label{sec:affine}
The affine Jensen--Mercer family [Appendix~\ref{app:affine}] interpolates the two bounds in one
parameter, but its extremal members carry all the sharpness.  For $f=y^{2}$ the tightest
upper bound is always the endpoint $c=-1$ (Mercer) and the tightest lower bound the
endpoint $c=1$ (Jensen).  
The extremization over both parameters is exact rather
than asymptotic. For fixed $c<0$ the bound is minimal at an endpoint of the admissible
$b$-interval and increases with $c$, so only $c=-1$ survives.  The upper endpoint is the
Mercer-type ceiling of
Eq.~(\ref{eq:mercertype}). 
No intermediate member improves on the two-sided sandwich of
Eq.~(\ref{eq:sandwich}), so the family interpolates between two bounds without adding
information to either.

%%%%%%%%%%%%%%%%%%%%%%%%%%%%%%%%%%%%
\subsection{The realizability gap}\label{sec:dyn}
The ceilings discussed so far are distributional.  The dynamical datum leads to a
\emph{realizability gap} of $\mathcal I$,
\begin{equation}
\Delta_{\mathrm{real}}(\mathcal I)\ \equiv\ U^\ast(\mathcal I)-\dot\sigma .
\label{eq:realgap}
\end{equation}

Consider $y=J/p$ on a ring
$[0,2\pi)$ with constant $D$ and steady flux $J>0$. In this case, the stationary Fokker--Planck equation is
equivalent to the Riccati equation
\begin{equation}
D\,y'=y^{2}-\mu f\,y ,
\label{eq:riccati}
\end{equation}
whose solutions are in one-to-one correspondence with the drives that produce them.  Integrating
\eqref{eq:riccati} over the ring gives two exact functionals of a periodic solution, the
normalization $(2\pi)^{-1}\int dx/y=1/v$ and
\begin{equation}
\frac{1}{2\pi}\int_0^{2\pi}y\,dx=\mu\bar f,\qquad
\bar f=\frac{1}{2\pi}\int_0^{2\pi}f\,dx.
\label{eq:arithmean}
\end{equation}
Together with $\dot\sigma=D^{-1}\langle y^{2}\rangle$ they fix the dissipation in terms of the
flux and the mean drive,
\begin{equation}
\dot\sigma=\frac{\mu\bar f\,v}{D}.
\label{eq:EPflux}
\end{equation}

For every steady current with $\min y=\alpha$ and $\max y=\beta$ one then has [Appendix~\ref{app:dyn}]
\begin{equation}
\dot\sigma=U_{\mathrm{BD}}-\frac{\mathrm{Var}_{\max}(y)-\mathrm{Var}(y)}{D}\ \le\ U_{\mathrm{BD}},
\label{eq:propdyn}
\end{equation}
with equality if and only if $y$ is constant, so every nontrivial steady state leaves the
realizability gap
\begin{equation}
\Delta_{\mathrm{real}}=\ U_{\mathrm{BD}}-\dot\sigma
=\frac{\mathrm{Var}_{\max}(y)-\mathrm{Var}(y)}{D}.
\label{eq:shapedeficit}
\end{equation}
The supremum over all smooth drives still equals $U_{\mathrm{BD}}$, since a two-level current with
$y=\beta$ on the fraction $f_\beta=\beta(v-\alpha)/[v(\beta-\alpha)]$ of the ring and $y=\alpha$
elsewhere attains it in the limit of smoothed jumps.  Stationarity therefore does not lower the
ceiling but forbids its attainment.  
A bound on the drive does lower the ceiling within the class of bounded drives.  Bounding the
slope of the current by the drive
interval gives
\begin{equation}
\dot\sigma\ \le\ U_{\mathrm{BD}}-\frac{v\,\mathcal
J(\alpha,\beta)}{\pi\,\kappa}
\ \le\ U_{\mathrm{BD}}-\frac{v\,\mathcal J(\alpha,\beta)}{\pi\,\bar\kappa}\ <\ U_{\mathrm{BD}},
\label{eq:boundeddrive}
\end{equation}
where $\mathcal J(\alpha,\beta)=\int_\alpha^\beta(y-\alpha)(\beta-y)\,dy/y$ is positive,
$\kappa=\max|y^{2}-\mu fy|$, and $\bar\kappa$ is the maximum compatible with the drive bound.
For a symmetric constraint $|f|\le F_{\max}$ one has $\bar\kappa=\beta(\beta+\mu F_{\max})$
exactly, and since $\mathcal J\ge w^{3}/(6\beta)$ with $w=\beta-\alpha$,
\begin{equation}
U_{\mathrm{BD}}-\dot\sigma\ \ge\ \frac{v\,w^{3}}{6\pi\beta\,\bar\kappa}\ \ge\ 
\frac{v\,w^{3}}{6\pi\beta^{2}(\beta+\mu F_{\max})},
\label{eq:narrowband}
\end{equation}
so the improvement is cubic in the current range and independent of $D$ once the data are
expressed through $v$, $\beta$ and $w$ [Appendix~\ref{app:dyn}].

\section{Variants}\label{5}
\subsection{Position-dependent diffusion}\label{sec:posdep}
For $D_i=D_i(x)$ the Langevin equation is read in the Stratonovich convention with drift
$b_{\mathrm S,i}=\mu_i f_i+\tfrac12\partial_iD_i$,
\begin{equation}
\dot x_i=b_{\mathrm S,i}(x)+\sqrt{2D_i(x)}\circ\eta_i ,
\label{eq:SDEpos}
\end{equation}
which keeps the Fokker--Planck flux in its thermodynamically natural
form $J_i=\mu_if_ip-D_i\partial_ip$ \cite{Risken,Kampen,Lau07}.  The bounds below are
formulated in terms of the physical flux $J_i$ and are therefore convention independent.

Averaging the pointwise inequality $y_i^{2}\le(\alpha_i+\beta_i)y_i-\alpha_i\beta_i$ with
weight $D_i^{-1}$ gives the ceiling
\begin{equation}
\dot\sigma_i\ \le\ \big\langle D_i^{-1}\big\rangle
\Big[(\alpha_i+\beta_i)\,\bar v_i^{(w)}-\alpha_i\beta_i\Big],\label{eq:posdep}
\end{equation}
in which 
\begin{equation}
\bar v_i^{(w)}=\frac{\langle y_i/D_i\rangle}{\langle 1/D_i\rangle}\in[\alpha_i,\beta_i].
\end{equation}
  For constant $D_i$,
Eq.~(\ref{eq:posdep}) reduces to Eq.~(\ref{eq:upper}).

On the opposite side the floor acquires the harmonic-type denominator
\begin{equation}
\dot\sigma_i=\Big\langle\frac{y_i^{2}}{D_i}\Big\rangle
\ \ge\ \frac{\langle y_i\rangle^{2}}{\langle D_i\rangle},
\label{eq:posdepfloor}
\end{equation}
by the Cauchy--Schwarz inequality
$\langle y_i\rangle^{2}\le\langle y_i^{2}/D_i\rangle\langle D_i\rangle$.  The pair
(\ref{eq:posdep}) and (\ref{eq:posdepfloor}) is the position-dependent counterpart of the
sandwich (\ref{eq:sandwich}), with $\langle D_i^{-1}\rangle$ and $\langle D_i\rangle$
replacing $D_i^{-1}$ and $D_i$ on the two sides.

For weak disorder, $D_i=D_i^{(0)}(1+\epsilon\delta_i)$ with $\langle\delta_i\rangle=0$ and
$\epsilon\ll1$, one has
$\langle D_i^{-1}\rangle=(D_i^{(0)})^{-1}[1+\epsilon^{2}\langle\delta_i^{2}\rangle
+O(\epsilon^{3})]$ and
\begin{equation}
\bar v_i^{(w)}=v_i-\epsilon\langle y_i\delta_i\rangle
+\epsilon^{2}\big[\langle y_i\delta_i^{2}\rangle-v_i\langle\delta_i^{2}\rangle\big]
+O(\epsilon^{3}),
\label{eq:vbarweak}
\end{equation}
the second-order term following from expanding numerator and denominator of
$\bar v^{(w)}=\langle y/D\rangle/\langle1/D\rangle$ to the same order.  Inserting
\eqref{eq:vbarweak} into Eq.~(\ref{eq:posdep}) gives the ceiling
\begin{align}
\dot\sigma_i\le\frac{1}{D_i^{(0)}}\Big[&(\alpha_i+\beta_i)v_i-\alpha_i\beta_i
-\epsilon(\alpha_i+\beta_i)\langle y_i\delta_i\rangle\nonumber\\
&+\epsilon^{2}\big((\alpha_i+\beta_i)\langle y_i\delta_i^{2}\rangle
-\alpha_i\beta_i\langle\delta_i^{2}\rangle\big)
+O(\epsilon^{3})\Big].
\label{eq:posdepweak}
\end{align}
 The floor,
by contrast, is exact. $\langle D_i\rangle=D_i^{(0)}(1+\epsilon\langle\delta_i\rangle)=D_i^{(0)}$
whenever $\langle\delta_i\rangle=0$, so that $v_i^{2}/\langle D_i\rangle=v_i^{2}/D_i^{(0)}$ to
all orders in $\epsilon$.  The sandwich therefore responds at first order to the spatial
correlation between the local current and the diffusivity, which is the mechanism by which a
position-dependent $D$ can shift the bounds at fixed mean drift.

%%%%%%%%%%%%%%%%%%%%%%%%%%%%%%%%%%%%
\subsection{Underdamped dynamics}\label{sec:underdamped}
For underdamped Langevin dynamics $\dot x_i=v_i$,
$m_i\dot v_i=f_i-\zeta_iv_i+\sqrt{2k_BT_i\zeta_i}\,\eta_i$, the density $p(x,v,t)$ obeys the
Kramers equation \cite{Risken,Kampen},
\begin{equation}
\partial_tp=-v_i\partial_{x_i}p
+\frac{1}{m_i}\partial_{v_i}\big[(\zeta_iv_i-f_i)p\big]
+\frac{k_BT_i\zeta_i}{m_i^{2}}\,\partial_{v_i}^{2}p ,
\label{eq:kramers}
\end{equation}
with the coordinate index $i$ summed over.  The entropy production rate is a single quadratic
form in the irreversible-current field \cite{LS24},
\begin{equation}
\dot\sigma=\sum_i D_i^{-1}\big\langle u_i^{2}\big\rangle,\qquad
u_i=v_i+\frac{k_BT_i}{m_i}\partial_{v_i}\ln p,
\label{eq:under}
\end{equation}
with $\langle u_i\rangle=\langle v_i\rangle$.  The field $u_i$ is the velocity measured
relative to the local drift of the density in velocity space.

Writing $\nu_i(x)=\int v_i\,p(v_i|x)\,dv_i$ for the conditional mean velocity, one has
$\mathbb E[u_i|x]=\nu_i(x)$ pointwise in $x$, since the $\partial_{v_i}\ln p$ term integrates
to a vanishing boundary term.  The law of total variance then gives the exact decomposition
\begin{align}
\dot\sigma&=\dot\sigma_{\mathrm{marg}}+\sum_i D_i^{-1}\mathbb E\big[(u_i-\nu_i)^{2}\big],\label{eq:exactdecomp}\\
\text{with}&~\dot\sigma_{\mathrm{marg}}=\sum_i D_i^{-1}\mathbb E\big[\nu_i^{2}\big],
\end{align}
whose second term is strictly nonnegative and vanishes if and only if the conditional
velocity distribution is the shifted Maxwellian of temperature $T_i$ with mean $\nu_i(x)$
[See Appendix~\ref{app:under}].  The decomposition therefore isolates exactly the part of the
dissipation that the marginal position current cannot account for.

Equation~(\ref{eq:under}) has the structure of Eq.~(\ref{eq:EP}) with $y\to u$, so whenever
$u_i\in[\alpha_i^{u},\beta_i^{u}]$, we have
\begin{equation}
D_i^{-1}\langle v_i\rangle^{2}\ \le\ \dot\sigma_i\ \le\
D_i^{-1}\big[(\alpha_i^{u}+\beta_i^{u})\langle v_i\rangle-\alpha_i^{u}\beta_i^{u}\big].
\label{eq:underbnd}
\end{equation}
Since $p=e^{-m_iv_i^{2}/2k_BT_i}\Psi$ gives $u_i=\tfrac{k_BT_i}{m_i}\partial_{v_i}\ln\Psi$, the
ceiling is finite if and only if the log-derivative of the nonthermal factor is bounded.

In the adiabatic regime of large friction the bounds survive, but the three quantities that
enter must be distinguished. The deficit of the local velocity variance with respect to its
thermal value, the excess dissipation carried by the non-Maxwellian part of the velocity
distribution, and the relative error of the entropy production rate itself,  are not of
the same order.  With $n(x)=\int p\,dv$ the position density,
$\nu(x)=\int v\,p(v|x)\,dv$ the conditional mean velocity,
$S(x)=\langle(v-\nu)^{2}|x\rangle$ the local velocity variance, and $\theta=k_BT/m$ its thermal
value, one has
\begin{equation}
S-\theta=\frac{k_BT}{\zeta}\,\nu\,\partial_x\ln n+O(\zeta^{-3}),
\label{eq:Sdev}
\end{equation}
in which the explicit term is $O(\zeta^{-2})$ since $\nu=(\zeta^{-1})(f-k_BT\partial_x\ln n)+
O(\zeta^{-3})$. The deficit of the local velocity variance is therefore second order in
$1/\zeta$.  The excess dissipation, by contrast, is a
fourth-order absolute quantity [Eq.~(\ref{eq:Ausq})], so that its ratio to the marginal term
is of order $\zeta^{-2}$.  The Hermite expansion of the velocity distribution around the local
Maxwellian gives
\begin{equation}
\dot\sigma_u=\dot\sigma_{\mathrm{marg}}\Big[1+O(\varepsilon^{2})\Big],
\label{eq:adiabatic}
\end{equation}
with $\varepsilon^{2}=mk_BT/(\zeta^{2}L^{2})=\tau_p/\tau_D$.  The
correction is nonnegative relative to $\dot\sigma_{\mathrm{marg}}$, since that is the
content of Eq.~(\ref{eq:exactdecomp}).  The order
estimate itself is a conditional adiabatic theorem, valid under the hypotheses listed in
Appendix~\ref{app:under}.

%%%%%%%%%%%%%%%%%%%%%%%%%%%%%%%%%%%%
\subsection{Tilted periodic potential}\label{sec:tilted}
For a single overdamped coordinate on a ring $[0,2\pi)$ with $V=V_0\cos x$, a tilt $F$, and
constant $D=k_BT/\zeta$, the entropy production rate is [Appendix~\ref{app:tilted}]
\begin{equation}
\dot\sigma=\frac{D\beta^{2}F^{2}}{M_1},
\label{eq:tiltedexact}
\end{equation}
with
\begin{equation}
M_1=\frac{1}{2\pi}\int_0^{2\pi}\frac{dx}{1+c(x)},\quad p=\frac{1}{2\pi}(1+c),
\end{equation}
For $g=\beta V_0$ and $a=\beta F$, we can expand
\begin{equation}
\begin{aligned}
M_1&=1+\frac{g^{2}}{2(1+a^{2})}+\frac{3\,g^{4}}{8(1+a^{2})(a^{2}+4)}+O(g^{6}),\\
\dot\sigma&=D\beta^{2}F^{2}\Big[1-\frac{g^{2}}{2(1+a^{2})}
+\tfrac{(5-a^{2})g^{4}}{8(1+a^{2})^{2}(a^{2}+4)}+O(g^{6})\Big].
\end{aligned}
\label{eq:sigexp}
\end{equation}
To leading order the bounds form the ladder
\begin{equation}
\underbrace{-1}_{\text{Jensen}}<\underbrace{-\tfrac12}_{\dot\sigma}
<\underbrace{0}_{\text{Bhatia--Davis}}<\underbrace{+1}_{\text{Mercer-type}},
\end{equation}
in units of $g^{2}/(1+a^{2})$ normalized to $D\beta^{2}F^{2}$.

The tightness of the upper bound is measured by the ratio
of the variance of $1/p$ to its maximal admissible value,
\begin{equation}
\rho\equiv\frac{\mathrm{Var}(1/p)}{\mathrm{Var}_{\max}(1/p)}
=\frac12+\rho_2\,g^{2}+\rho_4\,g^{4}+O(g^{6}),
\label{eq:rho}
\end{equation}
with
\begin{equation}
\rho_2=\frac{m_4-q_4/2}{\lambda^{2}},
\label{eq:rho2}
\end{equation}
where $\mathrm{Var}_{\max}(1/p)=(u_{\max}-2\pi)(2\pi-u_{\min})$, $\lambda=(1+a^{2})^{-1/2}$,
$m_4=3/[8(1+a^{2})(a^{2}+4)]$ is the $g^{4}$ coefficient of $M_1-1$, and $q_4$ is given by
Eq.~(\ref{eq:q4}) [cf. Appendix~\ref{app:tilted}].  In terms of $s=a^{2}$, we write
\begin{equation}
\rho_2=\frac{7s^{2}-9s-48}{8(1+s)(s+4)^{2}(s+9)}.
\end{equation}
The numerator is a quadratic in $s$ whose coefficients have a single sign change, so
$\rho_2$ has exactly one positive root,
$s_\ast=(9+\sqrt{1425})/14\simeq3.3392$, that is, $a_\ast\simeq1.8274$.  It is negative for
$0\le a<a_\ast$, positive for $a>a_\ast$, and decays as $\rho_2\sim7/(8a^{4})$ at large
tilt.  At weak barrier and moderate tilt the ceiling therefore \emph{loosens} relative to
the flat-tilt value $\rho=\tfrac12$.  

The strong-barrier regime ($g\gg1$ at fixed $a>0$) is exactly solvable to leading order
[cf. Appendix~\ref{app:tilted}].  The density is exponentially peaked at  $x=\pi$,
where $u=1/p$ attains its minimum $u_{\min}=\sqrt{2\pi/g}\,[1+O(1/g)]$, while the minimum of
$p$ lies in an interior boundary layer downwind of the barrier top,
\begin{equation}
x^{*}=\xi^{*}(a)\sqrt{2/g}+O(1/g),
\end{equation}
with $\xi^{*}$ the positive root of
$
\sqrt{\pi}\,\xi\,e^{\xi^{2}}\big[\coth(\pi a)-\mathrm{erf}(\xi)\big]=1$.
 The resulting maximum of $u$,
\begin{equation}
u_{\max}=\sqrt{\frac{2\pi^{2}}{g}}\;
\frac{e^{2g}}{(e^{a\pi}-e^{-a\pi})\,\gamma_1(a)}\,[1+O(g^{-1/2})],
\label{eq:umax}
\end{equation}
with $\gamma_1(a)\equiv1/[2\xi^{*}(a)]$, gives the relative gap
\begin{equation}
\begin{split}
\frac{U_{\mathrm{BD}}-\dot\sigma}{\dot\sigma}
&=\frac{4\pi^{2}+(2\pi-u_{\min})(u_{\max}-2\pi)}{4\pi^{2}M_1}-1\\
&\approx\frac{\sqrt{g}-1}{\sqrt2\,a\,\gamma_1(a)}
\ \xrightarrow[g\to\infty]{}\ \infty
\end{split}
\label{eq:stronggap}
\end{equation}
and the vanishing tightness ratio
\begin{equation}
\rho\ \approx\ \sqrt{\frac{2}{g}}\;a\,\gamma_1(a)\ \xrightarrow[g\to\infty]{}\ 0 .
\label{eq:strongrho}
\end{equation}
The ceiling is therefore \emph{least} tight in the sharply peaked regime, its gap growing as
$\sqrt g$ while the Jensen floor decays as $e^{-2g}$. The TUR floor is the one that survives,
with $L_{\mathrm{TUR}}/\dot\sigma\to\tanh(\pi a)/(\pi a)$.

\begin{figure*}[t]
\centering
\includegraphics[width=\textwidth]{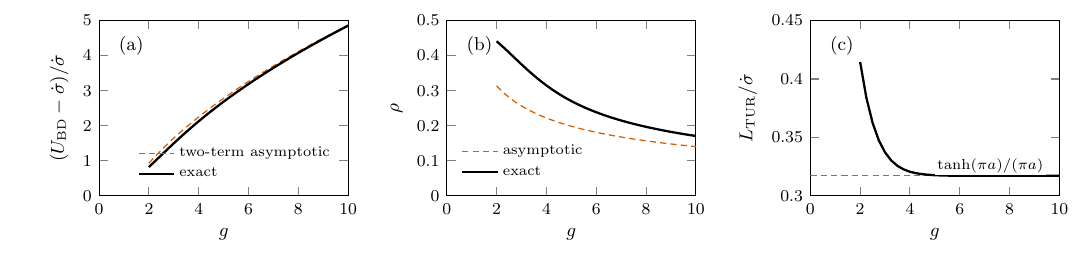}
\caption{Strong-barrier regime at $a=1$, from the exact solution.  (a) Relative gap of
the Bhatia--Davis ceiling, Eq.~(\ref{eq:stronggap}).  It grows as $\sqrt g$, shown dashed
for the two-term asymptotic.  (b) Tightness ratio $\rho$ of Eq.~(\ref{eq:rho}).  It decays
as $g^{-1/2}$, shown dashed for the leading asymptotic $\sqrt{2/g}\,a\gamma_1$.  (c) The TUR
floor relative to the exact rate.  It tends to $\tanh(\pi a)/(\pi a)$, shown dashed, so the
TUR survives the barrier while the Jensen floor, $\propto e^{-2g}$, does not.}
\label{fig:strong}
\end{figure*}

The leading-order asymptotics converge slowly, since the saddle points are shifted by the
tilt and the corrections are $O(g^{-1/2})$, so the scaling laws are best verified numerically
(Fig.~\ref{fig:strong}).  At $a=1$ the exact evaluation gives
$(U_{\mathrm{BD}}-\dot\sigma)/\dot\sigma=3.19$ at $g=6$ and $4.08$ at $g=8$, against $3.26$
and $4.12$ from the two-term formula. The tightness ratio is $0.239$ and $0.197$, and
$\rho\sqrt g$ decreases toward $\sqrt2\,a\gamma_1(1)\simeq0.44$ from above.  At $a=1$ and
$g=6$ the TUR ratio is $0.3172$, against $\tanh\pi/\pi=0.3171$.  These asymptotics hold at
fixed $a>0$.  The equilibrium case $a=0$ is degenerate, since $J\to0$, and must be taken
separately. The extrema of $u$ then revert to the barrier top $x=0$, with
$u_{\max}=2\pi I_0(g)e^{g}$ and $u_{\min}=2\pi I_0(g)e^{-g}$, where $I_0$ is the modified
Bessel function, so that $\rho\to(2\pi g)^{-1/2}$.

When the tilt is known, Eq.~(\ref{eq:EPflux}) fixes $\dot\sigma$ exactly, so the two-sided
bound describes the shape of the current rather than inferring its dissipation.  For the data
$\{v,\alpha,\beta\}$ the tightness ratio is then a functional of the range alone,
\begin{equation}
\rho=\frac{v\,(\mu F-v)}{(\beta-v)(v-\alpha)}=\frac{\mu F-v}{\hat a-v},\qquad
\hat a\equiv(\alpha+\beta)-\frac{\alpha\beta}{v},
\label{eq:rhoTilt}
\end{equation}
and the data are LP-tight exactly when $\mu F=\hat a$.  The fixed-tilt family is nonempty for
every $0<v<\mu F$ and approaches $U_{\mathrm{BD}}$ as its supremum, which no smooth member
attains [Appendix~\ref{app:dyn}].

\section{Example: The Brownian motor}\label{sec:model}
To illustrate the bounds on a concrete physical system, we take the tilted periodic
(``washboard'') potential \cite{Risken}, a standard model of a Brownian motor
\cite{Reimann02,Haenggi09} and of a molecular motor moving along its track
\cite{Julicher97,Kolomeisky07}.  Working in units in which the potential period is $2\pi$,
energies
are measured in $k_BT$, and times in the diffusive time over one period, the model is
specified by the two dimensionless control parameters
\begin{equation}
a\equiv \beta F,\qquad g\equiv \beta V_0 ,
\end{equation}
where $F$ is the tilt (the driving force) and $V_0$ the barrier amplitude.  The tilt alone
would give the uniform-drift rate $\dot\sigma=D\beta^{2}F^{2}$.  The periodic corrugation
reduces both the mean velocity and the dissipation.

Using the exact relation of Sec.~\ref{sec:tilted} with
$M_1=1+\tfrac{g^{2}}{2(1+a^{2})}+\tfrac{3g^{4}}{8(1+a^{2})(a^{2}+4)}+O(g^{6})$, together
with the Jensen floor, the Bhatia--Davis ceiling, the Mercer-type ceiling, and the TUR floor
$L_{\mathrm{TUR}}=v^{2}/D_{\mathrm{eff}}$ of Eq.~(\ref{eq:turfloor}), the numbers in
Table~\ref{tab:model} are obtained for three representative parameter sets.  All entries are
in units of $D\beta^{2}F^{2}$.  The values are evaluated from the exact Fokker--Planck
solution of Appendix~\ref{app:tilted}, and the $O(g^{4})$ expansions of
Eqs.~(\ref{eq:sigexp}), (\ref{eq:q4}) and~(\ref{eq:Adeffexp}) reproduce them to a relative
accuracy of $5\cdot10^{-5}$ for the rate and $1.2\cdot10^{-3}$ for the ceiling when
$g\le0.5$, and to $2.6\cdot10^{-3}$ and $1.7\cdot10^{-2}$ at $g=1$.\footnote{The Jensen and exact-rate columns use
$\dot\sigma=D\beta^{2}F^{2}/M_1$ and Eq.~(\ref{eq:sigexp}).  The Bhatia--Davis and Mercer-type
columns use the exact $\mathrm{Var}_{\max}(1/p)$, obtained from the numerically determined
extrema of $1/p$, whose $O(g^{4})$ form is Eq.~(\ref{eq:q4}). The Mercer-type column exceeds the
Bhatia--Davis one by $\mathrm{Var}_{\max}/D$.  The TUR column uses
Eqs.~(\ref{eq:Adeff}) and (\ref{eq:Adeffexp}).}

\begin{table}[t]
\caption{Bounds on the entropy production rate of the tilted periodic (Brownian-motor)
model for three parameter sets $(a,g)=(\beta F,\beta V_0)$.  All entries are normalized to
$D\beta^{2}F^{2}$.  The ``exact'' column uses $\dot\sigma=D\beta^{2}F^{2}/M_1$, and the TUR
column uses Eq.~(\ref{eq:Adeffexp}).  The exact entries come from the converged mode sum of
Appendix~\ref{app:tilted}, with all quoted digits stable under further truncation.  The last column is the
tightness ratio $\rho$ of Eq.~(\ref{eq:rho}).  BD labels the Bhatia--Davis ceiling and
Mercer the Mercer-type ceiling.}
\small
\begin{ruledtabular}
\begin{tabular}{ccccccc}
$(a,g)$ & Jensen & exact $\dot\sigma$ & BD & Mercer & TUR & $\rho$\\
\hline
$(1.0,\,0.3)$ & $0.9559$ & $0.9777$ & $0.9996$ & $1.0433$ & $0.9358$ & $0.4989$\\
$(1.0,\,0.5)$ & $0.8818$ & $0.9391$ & $0.9970$ & $1.1122$ & $0.8360$ & $0.4968$\\
$(2.0,\,0.5)$ & $0.9507$ & $0.9750$ & $0.9994$ & $1.0480$ & $0.9016$ & $0.5002$\\
\end{tabular}
\end{ruledtabular}
\label{tab:model}
\end{table}

\begin{figure*}[t]
\centering
\includegraphics[width=\textwidth]{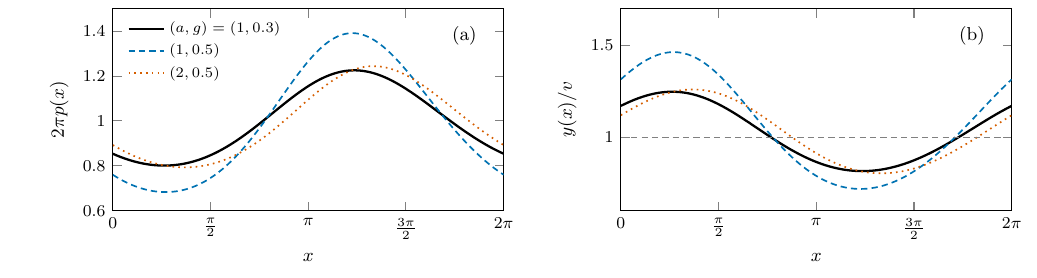}
\caption{Stationary state of the tilted periodic (Brownian-motor) model for the
three parameter sets of Table~\ref{tab:model}, from the exact Fokker--Planck
solution.  (a) Stationary density $p(x)$, scaled as $2\pi p$ (flat-tilt value unity),
maximal at the well bottom $x=\pi$ and minimal near the barrier top $x=0$.  (b) Local
current velocity $y(x)=J/p(x)$ in units of its mean $v=\langle y\rangle$ (dashed line),
whose spread about unity is the fluctuation exploited by the ceiling of
Eq.~(\ref{eq:upper}) and whose extrema are $\alpha/v$ and $\beta/v$.}
\label{fig:model}
\end{figure*}

\begin{figure*}[t]
\centering
\includegraphics[width=\textwidth]{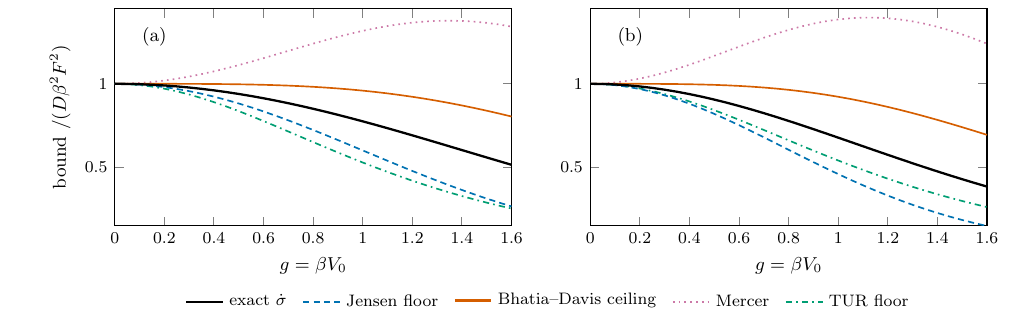}
\caption{Bounds on $\dot\sigma$ as functions of the barrier height $g=\beta V_0$ at
fixed tilt, in units of $D\beta^2F^2$.  Shown are the exact rate (black), the Jensen floor
(blue dashed), the Bhatia--Davis ceiling (red), the Mercer bound (purple dotted), and the
TUR floor (green dash-dotted), all evaluated from the exact Fokker--Planck solution.  (a)
$a=1$, where $c_2>0$ and hence $D_{\mathrm{eff}}>D$, so the TUR floor lies below the Jensen
floor.  (b) $a=0.5<1/\sqrt3$, where $c_2<0$ and the ordering of the two floors is reversed
[Eq.~(\ref{eq:Adeffexp})].  In both panels the exact rate stays inside the two-sided
sandwich of Eq.~(\ref{eq:sandwich}), and the Mercer bound lies above the ceiling by
$\mathrm{Var}_{\max}(y)/D$.}
\label{fig:bounds}
\end{figure*}

Figure~\ref{fig:model} shows the stationary density and the local current velocity whose
extrema and mean fix the ceilings of Table~\ref{tab:model}, and
Fig.~\ref{fig:bounds} shows how the bounds vary with the barrier height at fixed tilt.  The
relative gap of the Bhatia--Davis ceiling is at most $6.2\%$, and the ceiling stays within $0.3\%$
of the no-barrier value $D\beta^{2}F^{2}$ because the $O(g^{2})$ corrections cancel identically,
\begin{align}
\frac{U_{\mathrm{BD}}}{D\beta^{2}F^{2}}
&=\frac{1+\mathrm{Var}_{\max}(u)/(2\pi)^{2}}{M_1^{2}}\nonumber\\
&=1-\frac{s^{3}+24s^{2}+79s+96}{4(1+s)^{2}(s+4)^{2}(s+9)}\,g^{4}+O(g^{6}),
\label{eq:BDweak}
\end{align}
with $s=a^{2}$ and $u=1/p$.  The cancellation is between the drop of $v^{2}$ and the growth of
$\mathrm{Var}_{\max}$ at $O(g^{2})$, and it uses
$2\pi(u_{\min}+u_{\max})-u_{\min}u_{\max}=4\pi^{2}+(2\pi-u_{\min})(u_{\max}-2\pi)$ together with
$m_2=\lambda^{2}/2$.  The first nonzero correction is of order $g^{4}$ and negative.  At the
tabulated parameters Eq.~(\ref{eq:BDweak}) gives $0.99960$ and $0.99935$ for $(a,g)=(1,0.3)$ and
$(2,0.5)$, against $0.9996$ and $0.9994$ in Table~\ref{tab:model}.  The sharp ceiling therefore
says little at weak barrier and acquires content at strong barrier, where the gap grows as
$\sqrt{g}$, and with a bounded drive (Sec.~\ref{sec:dyn}).  The Jensen floor, by contrast,
drops markedly with $g$, the tightness ratio decreasing from $\rho=0.4989$ to $0.4968$ for $a=1$
as $g$ goes from $0.3$ to $0.5$, consistent with $\rho\to0$.  All entries except the Mercer bound
lie below the uniform-drift value, while the Mercer bound exceeds the ceiling by exactly
$\mathrm{Var}_{\max}(y)/D$.  At these tilts $c_2>0$ and the Jensen floor dominates the TUR floor,
which falls short of the rate by the relative gap $2a^{2}g^{2}/(1+a^{2})^{2}$.

As a physical calibration, for a molecular motor \cite{Kolomeisky07,Julicher97} with track
period $L$ one has
$a=\beta FL/(2\pi)$ and $g=\beta V_0$.  Taking $L=8$~nm, $k_BT=4.14$~pN$\,$nm, and a stall
force $F\simeq5$~pN gives $a\simeq1.5$, while $V_0\simeq0.3$--$0.5\,k_BT$ corresponds to a
weakly corrugated landscape.  The estimate $\dot\sigma\simeq0.98\,D\beta^{2}F^{2}$ for
$a=1$ indicates that the dissipation of a weakly corrugated motor is bounded above by the
uniform-drift value to within a few percent, which the Jensen floor alone cannot provide.  A
systematic exploration at larger $g$, where the TUR floor is the surviving bound, and the
collective-transport model of Ref.~\cite{LS22} are left for future work.

%%%%%%%%%%%%%%%%%%%%%%%%%%%%%%%%%%%%
\section{Summary and discussion}\label{sec:summary}
%%%%%%%%%%%%%%%%%%%%%%%%%%%%%%%%%%%%
The convexity underlying the Jensen bound also yields a complementary upper bound.  For the
quadratic functional of overdamped Langevin dynamics, Mercer's reflection inequality and the
Bhatia--Davis chord inequality are the same pointwise statement.  Averaging the chord form gives
the sharp ceiling, which is the convex dual of the smallest affine majorant of the
entropy-production quadratic over the current range, while averaging the reflection form gives a
Mercer-type ceiling larger by one maximal variance,.  The resulting sandwich, Eq.~(\ref{eq:sandwich}), involves only the mean
drift and the extremal local current, and it extends to arbitrary diffusion tensors, to
position-dependent diffusion and to underdamped dynamics.

The bounds we have derived depend, however, on layers of  assumptions.  The sharp sandwich needs the local current field to be bounded and the averages defining $v$ to exist.  The position-dependent formulation is stated in the Stratonovich convention and its floor
uses $\langle\delta\rangle=0$.  The underdamped ceiling needs the field $u$ to be bounded, and the
adiabatic statement is conditional on the three hypotheses of Appendix~\ref{app:under}.
In spite of  these assumptions,  the present study is still useful in broad topics of theory, simulation, and experiment.

%%%%%%%%%%%%%%%%%%%%%%%%%%%%%%%%%%%%%%%%%%%%%%%%%%%%%%%%%%%%%%
\begin{acknowledgments}
X.G. is supported by Yancheng Institute of Technology (xjr2024030). Z.H. is supported by the National Natural Science Foundation of China under Grant No.
12305035.
\end{acknowledgments}

%%%%%%%%%%%%%%%%%%%%%%%%%%%%%%%%%%%%%%%%%%%%%%%%%%%%%%%%%%%%%%
\appendix
\renewcommand{\thesubsection}{\thesection.\arabic{subsection}}
\section{Affine-majorant and LP duality}\label{app:LP}

This appendix proves the dual representation of the sharp ceiling, Eq.~(\ref{eq:dual}), and
establishes the properties of the value function $L^\ast(v)$ used in Sec.~\ref{sec:general}.

\subsection{Primal problem}  Let $\mathcal B=\prod_{i=1}^{N}[\alpha_i,\beta_i]$ be the box and
$Q(y)=y\!\cdot\!D^{-1}y$, with $D^{-1}$ positive semidefinite.  For $v\in\mathcal B$ the
sharp ceiling of Eq.~(\ref{eq:LP}) is
\begin{equation}
U^\ast(v)=\sup_{\mu\in\mathcal P(\mathcal B)}
\Big\{\int_{\mathcal B}Q\,d\mu\ :\ \int_{\mathcal B}y\,d\mu=v\Big\},
\label{eq:Aprimal}
\end{equation}
where $\mathcal P(\mathcal B)$ denotes the probability measures on $\mathcal B$.  Because
$\mathcal B$ is compact, $\mathcal P(\mathcal B)$ is weakly compact \cite{Billingsley} and
$\mu\mapsto\int y\,d\mu$ is weakly continuous, so the admissible set in
Eq.~(\ref{eq:Aprimal}) is weakly compact and the supremum is attained.  We write
$U^\ast(v)=\max$ in what follows.

\subsection{Affine-majorant lemma}  Let $L(y)=\mu_0+\lambda\!\cdot\!y$ be affine.  If
$L(c^{(s)})\ge Q(c^{(s)})$ at all $2^{N}$ vertices $c^{(s)}$ of $\mathcal B$, then $L\ge Q$
on $\mathcal B$.  Indeed, every $y\in\mathcal B$ is a convex combination
$y=\sum_{s}\theta_sc^{(s)}$ with $\theta_s\ge0$ and $\sum_s\theta_s=1$, so convexity of $Q$
gives
\begin{equation}
Q(y)\le\sum_s\theta_sQ(c^{(s)})\le\sum_s\theta_sL(c^{(s)})=L(y).
\label{eq:Amaj}
\end{equation}
Averaging Eq.~(\ref{eq:Amaj}) over an admissible $\mu$ and using $\int y\,d\mu=v$ yields
\begin{equation}
\int_{\mathcal B}Q\,d\mu\ \le\ \mu_0+\lambda\!\cdot\!v ,
\label{eq:Aweak}
\end{equation}
so every affine majorant of $Q$ bounds $U^\ast(v)$ from above.

\subsection{Dual program}  Equation~(\ref{eq:Aweak}) implies
\begin{equation}
U^\ast(v)\ \le\ L^\ast(v)\equiv\inf\{\,L(v)\ :\ L\ \text{affine},\ L\ge Q\
\text{on}\ \mathcal B\,\}.
\label{eq:Amajinf}
\end{equation}
By the lemma the constraint $L\ge Q$ on $\mathcal B$ is equivalent to its restriction to the
vertices, so Eq.~(\ref{eq:Amajinf}) is the linear program
\begin{equation}
L^\ast(v)=\min_{\mu_0,\lambda}\Big\{\mu_0+\lambda\!\cdot\!v\ :\
\mu_0+\lambda\!\cdot\!c^{(s)}\ge Q(c^{(s)}),\ s=1,\dots,2^{N}\Big\},
\label{eq:Adual}
\end{equation}
which is Eq.~(\ref{eq:dual}) of the main text.  Attaching the multipliers $p_s\ge0$ to the
$2^{N}$ constraints and using that $\mu_0$ and $\lambda$ are free variables, the dual of
Eq.~(\ref{eq:Adual}) is
\begin{equation}
\max_{p}\Big\{\sum_{s=1}^{2^{N}}p_sQ(c^{(s)})\ :\ p_s\ge0,\
\sum_sp_s=1,\ \sum_sp_sc^{(s)}=v\Big\},
\label{eq:Avertex}
\end{equation}
in which $\sum_sp_s=1$ and $\sum_sp_sc^{(s)}=v$ are the stationarity conditions with respect
to $\mu_0$ and $\lambda$.

\subsection{Strong duality}  Both programs are feasible.  In Eq.~(\ref{eq:Adual}) one may take
$\lambda=0$ and $\mu_0=\max_sQ(c^{(s)})$, while Eq.~(\ref{eq:Avertex}) admits the
barycentric weights of $v$ with respect to the vertices.  Both are bounded, the former from
below by $Q(v)$ according to Eq.~(\ref{eq:Amaj}) and the latter by the boundedness of $Q$ on
$\mathcal B$.  Hence, by the strong-duality theorem of linear programming \cite{Bertsimas},
both optima are attained and coincide,
\begin{equation}
U^\ast(v)=L^\ast(v)=\max_{p}\sum_{s=1}^{2^{N}}p_sQ(c^{(s)}).
\label{eq:Astrong}
\end{equation}
The equality $U^\ast=L^\ast$ combines weak duality, Eq.~(\ref{eq:Aweak}), with the
admissibility of the vertex measure that realizes the maximum in
Eq.~(\ref{eq:Avertex}).

\subsection{Support of the optimal measure}  Program~(\ref{eq:Avertex}) has $2^{N}$ variables and
$N+1$ equality constraints, so each basic feasible solution has at most $N+1$ nonvanishing
weights.  Equivalently, the optimum is attained at an extreme point of the admissible set,
supported on at most $N+1$ points of $\mathcal B$ by Carath\'eodory's theorem
\cite{Carath},
\begin{equation}
U^\ast(v)=\sum_{j=1}^{N+1}p_jQ(y^{(j)}),\qquad
\sum_{j=1}^{N+1}p_jy^{(j)}=v .
\label{eq:Acara}
\end{equation}

\subsection{Value function}  Equation~(\ref{eq:Avertex}) expresses $L^\ast(v)$ as a maximum over
convex combinations with barycenter $v$.  Convex combinations of the optimal weights for two
means $v$ and $v'$ give
\begin{equation}L^\ast(tv+(1-t)v')\ge tL^\ast(v)+(1-t)L^\ast(v')\end{equation} for $t\in[0,1]$, so $L^\ast$ is concave.
Since the feasible set of Eq.~(\ref{eq:Adual}) has finitely many constraints, the optimal
basis is constant on each cell of a polyhedral subdivision of $\mathcal B$, and $L^\ast$ is
piecewise linear.  It satisfies $L^\ast(c^{(s)})=Q(c^{(s)})$ at every vertex and is the
least concave majorant of $Q$ on $\mathcal B$.  Finally, if $\mathcal B'\subset\mathcal B$ the feasible set of
Eq.~(\ref{eq:Adual}) for $\mathcal B$ is contained in that for $\mathcal B'$, so
\begin{equation}L^\ast_{\mathcal B'}(v)\le L^\ast_{\mathcal B}(v).\end{equation} The ceiling tightens as the prior
information improves.

\subsection{Diagonal diffusion tensor}  For
$D^{-1}=\mathrm{diag}(D_1^{-1},\dots,D_N^{-1})$ the affine function
\begin{equation}
L(y)=\sum_{i=1}^{N}\frac{(\alpha_i+\beta_i)y_i-\alpha_i\beta_i}{D_i}
\label{eq:Adiag}
\end{equation}
majorizes $Q$, since each term is the chord majorant of $y_i^{2}/D_i$ on
$[\alpha_i,\beta_i]$.  At a vertex $y_i\in\{\alpha_i,\beta_i\}$ for every $i$, so each term
is tight and $L(c^{(s)})=Q(c^{(s)})$.  Hence $L$ is optimal in Eq.~(\ref{eq:Adual}) for all
$v\in\mathcal B$, and
\begin{equation}
L^\ast(v)=\sum_{i=1}^{N}\frac{(\alpha_i+\beta_i)v_i-\alpha_i\beta_i}{D_i},
\label{eq:AdiagL}
\end{equation}
which is Eq.~(\ref{eq:upper}) of the main text.

\subsection{Eigenmode majorant}  For a general positive-semidefinite $D^{-1}$ diagonalize
$D^{-1}=U\Lambda U^\top$ and set $z_k=u_k\!\cdot\!y$, so that $Q=\sum_k\lambda_kz_k^{2}$
and each mode ranges over $[\alpha^{(k)},\beta^{(k)}]$ with
$\alpha^{(k)}=\sum_i\min(u_{k,i}\alpha_i,u_{k,i}\beta_i)$ and
$\beta^{(k)}=\sum_i\max(u_{k,i}\alpha_i,u_{k,i}\beta_i)$.  The per-mode inequality
\begin{equation}
z_k^{2}\le(\alpha^{(k)}+\beta^{(k)})z_k-\alpha^{(k)}\beta^{(k)}
\label{eq:Aeigenb}
\end{equation}
holds pointwise, so the affine function
\begin{equation}
M(y)=\sum_k\lambda_k\big[(\alpha^{(k)}+\beta^{(k)})z_k-\alpha^{(k)}\beta^{(k)}\big]
\label{eq:AeigenM}
\end{equation}
majorizes $Q$ and reproduces Eq.~(\ref{eq:eigen}) of the main text.  It is admissible in
Eq.~(\ref{eq:Adual}) and therefore satisfies $M(v)\ge L^\ast(v)$ for all $v\in\mathcal B$,
with equality for diagonal $D$, in which case Eq.~(\ref{eq:AeigenM}) reduces to
Eq.~(\ref{eq:AdiagL}).

\subsection{Coordinated (four-corner) bound}  Since $Q$ is jointly convex, the coordinated
Jensen--Mercer lemma \cite{Toseef25} gives
$Q(\alpha+\beta-y)\le\sum_cQ(c)-Q(y)$ for $y\in\mathcal B$.  Averaging with Jensen's
inequality on the left then yields the weaker four-corner bound
\begin{equation}
\dot\sigma\le\sum_{c}Q(c)-Q(\alpha+\beta-v).
\label{eq:Afour}
\end{equation}

\subsection{Explicit $N=2$ illustration}  Take
$D^{-1}=\bigl(\begin{smallmatrix}2&1\\1&2\end{smallmatrix}\bigr)$ and
$\mathcal B=[0,1]^{2}$, so that $Q(y)=2y_1^{2}+2y_2^{2}+2y_1y_2$ and the four vertex values
are $Q=0,2,2,6$ at $(0,0)$, $(1,0)$, $(0,1)$, and $(1,1)$.

{Primal:}  A vertex measure is fixed by the weight $p_{11}$ of $(1,1)$, since the mean
and the normalization determine the remaining weights,
\begin{equation}
p_{10}=v_1-p_{11},\qquad p_{01}=v_2-p_{11},\qquad
p_{00}=1-v_1-v_2+p_{11},
\label{eq:AN2p}
\end{equation}
so that feasibility requires $\max(0,v_1+v_2-1)\le p_{11}\le\min(v_1,v_2)$.  The objective
\begin{equation}
\sum_sp_sQ(c^{(s)})=2p_{10}+2p_{01}+6p_{11}=2(v_1+v_2)+2p_{11}
\label{eq:AN2obj}
\end{equation}
increases with $p_{11}$, so the optimum is $p_{11}=\min(v_1,v_2)$ and
\begin{equation}
L^\ast(v)=2\big[v_1+v_2+\min(v_1,v_2)\big]
=\begin{cases}2v_1+4v_2, & v_1\ge v_2,\\ 4v_1+2v_2, & v_2\ge v_1,\end{cases}
\label{eq:AN2env}
\end{equation}which is
a concave piecewise-linear function whose two pieces are glued along the diagonal
$v_1=v_2$.  At $v=(\tfrac12,\tfrac12)$ the optimum is carried by the two-point measure
$\tfrac12\delta_{(0,0)}+\tfrac12\delta_{(1,1)}$, for which $\langle Q\rangle=3$.   The marginals
$(\tfrac12,\tfrac12)$ are also those of the product measure
$(\tfrac12\delta_0+\tfrac12\delta_1)^{\otimes2}$, and of the two-point measure $\tfrac12\delta_{(0,1)}+\tfrac12\delta_{(1,0)}$.
All three share the same marginals, and they give $\langle Q\rangle=3$, $\tfrac52$ and $2$
respectively.  Which coupling is extremal is decided by the program \eqref{eq:dual}, not by the
marginals. For a separable $\langle Q\rangle$ all three coincide, which is the diagonal case of
Sec.~\ref{sec:mercer}.

{Dual:}  The constraints of Eq.~(\ref{eq:Adual}) read $\mu_0\ge0$,
$\mu_0+\lambda_2\ge2$, $\mu_0+\lambda_1\ge2$, and $\mu_0+\lambda_1+\lambda_2\ge6$.  At
$v=(\tfrac12,\tfrac12)$ the last of them gives
\begin{equation}
\mu_0+\tfrac12(\lambda_1+\lambda_2)\ge\mu_0+\tfrac12(6-\mu_0)
=3+\tfrac12\mu_0\ge3,
\label{eq:AN2dual}
\end{equation}
with equality for $\mu_0=0$ and $\lambda=(3,3)$, i.e., for the affine majorant
$3(y_1+y_2)$.  The tight constraints are those of the vertices $(0,0)$ and $(1,1)$, which
are exactly the support of the optimal measure.  At $v=(1,0)$ it gives $3$ although the envelope
is $L^\ast=2$, attained instead by $2y_1+4y_2$.

The eigenvalues of $D^{-1}$ are $1$ and $3$, with
eigenvectors $u_{1,2}\propto(1,\pm1)/\sqrt2$, so that $z_{1,2}=(y_1\pm y_2)/\sqrt2$ range
over $[0,\sqrt2]$ and $[-1/\sqrt2,1/\sqrt2]$.  Equation~(\ref{eq:AeigenM}) then yields
$M(y)=3(y_1+y_2)+\tfrac12$, whose value at $v=(\tfrac12,\tfrac12)$ is $7/2$.  The four-corner
bound (A13) gives $10-Q(\tfrac12,\tfrac12)=17/2$, and the Jensen floor is
$v\!\cdot\!D^{-1}v=3/2$.  The bounds therefore order as
$\tfrac32\le\dot\sigma\le L^\ast=3\le M=\tfrac72\le\tfrac{17}{2}$.

\section{Optimal bounds from partial observations}\label{app:obs}
This appendix derives the closed forms quoted in Sec.~\ref{sec:obs}.  Throughout, $Q(y)=y^{2}/D$
and $\mu$ is a probability measure for the local current velocity.

\subsection{Moment problem}  For a constraint class $\mathcal C$ the ceiling of
Eq.~(\ref{eq:hier}) is
\begin{equation}
U^{\ast}(v;\mathcal C)=\sup_{\mu\in\mathcal C}\frac{1}{D}\int y^{2}\,d\mu ,
\label{eq:Bprimal}
\end{equation}
with the mean fixed, $\int y\,d\mu=v$.  Each class below is described by the normalization
and the mean together with at most one further affine constraint, so the feasible set is a
convex set of measures and the maximum of the linear functional $\int y^{2}d\mu$ is attained
at an extreme point.  A measure that is extreme for $m$ affine constraints in addition to
the normalization is supported on at most $m+1$ points \cite{Krein}, since on more points the
constraint functions would be affinely dependent and the weights could be perturbed without
leaving the
feasible set.  Hence the ceilings below are attained on two points, or on three when the
tail constraint is active, and are sharp rather than merely upper estimates.  The argument
is the one used in Appendix~\ref{app:LP}. The primal over measures is a linear program whose
extreme points are finitely supported.

\subsection{Mean only}  For \begin{equation}\mathcal C_\infty=\{\mu:\int y\,d\mu=v\}\end{equation} the functional is unbounded.  The
symmetric two-point measures
$\mu_R=\tfrac12\delta_{v-R}+\tfrac12\delta_{v+R}$, with $R>0$, have mean $v$ and second moment
$v^{2}+R^{2}$, which diverges as $R\to\infty$.  Hence
\begin{equation}
U^{\ast}=\infty .
\label{eq:Binf}
\end{equation}

\subsection{Mean and range}  For
\begin{equation}\mathcal C_\text{BD}=\{\mu:\operatorname{supp}\mu\subset[\alpha,\beta],\ \int y\,d\mu=v\}\end{equation} the
pointwise chord inequality $y^{2}\le(\alpha+\beta)y-\alpha\beta$ gives
\begin{equation}
U^{\ast}\le\frac{(\alpha+\beta)v-\alpha\beta}{D}\equiv U_{\mathrm{BD}} ,
\label{eq:BBD}
\end{equation}
and the two-point measure
\begin{equation}
\mu_{\mathrm{BD}}=\frac{v-\alpha}{\beta-\alpha}\,\delta_{\beta}
+\frac{\beta-v}{\beta-\alpha}\,\delta_{\alpha}
\label{eq:BBDmeas}
\end{equation}
attains it, since
$\int y^{2}d\mu_{\mathrm{BD}}=v^{2}+(v-\alpha)(\beta-v)=(\alpha+\beta)v-\alpha\beta$.
This is the Bhatia--Davis ceiling \cite{BD00} of Eq.~(\ref{eq:upper}), and
$\mu_{\mathrm{BD}}$ is the extreme point of the class.

\subsection{Mean and one-sided range}  For
\begin{equation}\mathcal C_1=\{\mu:\operatorname{supp}\mu\subset[0,\beta],\ \int y\,d\mu=v\}\end{equation} the
inequality $y^{2}\le\beta y$ holds on $[0,\beta]$, so
\begin{equation}
U_1=\frac{\beta v}{D},
\label{eq:Bone}
\end{equation}
attained by $\mu_1=[(\beta-v)/\beta]\,\delta_0+(v/\beta)\,\delta_\beta$.  For
$0\le\alpha\le v$ the inclusion $[\alpha,\beta]\subset[0,\beta]$ makes the one-sided class
the larger one, and
\begin{equation}
U_1-U_{\mathrm{BD}}=\frac{\alpha(\beta-v)}{D}\ge0
\label{eq:Bsingle}
\end{equation}
is the excess of the one-sided ceiling over the sharp one.  For $\alpha<0$ the inclusion is reversed,
$[0,\beta]\subset[\alpha,\beta]$, and the one-sided ceiling is the tighter of the two.

\subsection{Mean, range, and tail mass}  Let
\begin{equation}\mathcal C_\epsilon=\{\mu:\operatorname{supp}\mu\subset[\alpha,R],\
\mu((\beta,R])\le\epsilon,\ \int y\,d\mu=v\}\end{equation} with $\alpha\le v\le\beta\le R$, and let
$p=\mu((\beta,R])$ and $m_+$ be the mean of the part of $\mu$ supported on $(\beta,R]$.  On
$(\beta,R]$ the chord inequality $y^{2}\le(\beta+R)y-\beta R$ holds, and on $[\alpha,\beta]$
the corresponding inequality $y^{2}\le(\alpha+\beta)y-\alpha\beta$ holds.  Applying them to
the two parts of $\mu$ and using $\int y\,d\mu=v$ gives
\begin{align}
\int y^{2}d\mu\ &\le\ (\alpha+\beta)v-\alpha\beta+p(R-\alpha)(m_+-\beta)\nonumber\\
&\le\ (\alpha+\beta)v-\alpha\beta+p(R-\alpha)(R-\beta),
\label{eq:Btail}
\end{align}
where the second inequality uses $m_+\le R$.  Hence
\begin{equation}U^{\ast}\le U_{\mathrm{BD}}+\epsilon(R-\alpha)(R-\beta)/D.\end{equation}  Moreover,
$\mathcal C_\epsilon\subset\{\mu:\operatorname{supp}\mu\subset[\alpha,R]\}$, so the
two-sided ceiling on the enlarged range gives the second upper bound
\begin{equation}
U^{\ast}\le\frac{(\alpha+R)v-\alpha R}{D}
=U_{\mathrm{BD}}+\frac{(v-\alpha)(R-\beta)}{D},
\label{eq:Btail2}
\end{equation}
the two expressions being identical.  Combining Eqs.~(\ref{eq:Btail}) and
(\ref{eq:Btail2}),
\begin{equation}
U_\epsilon=U_{\mathrm{BD}}+\frac{1}{D}\min\{\epsilon(R-\alpha)(R-\beta),\
(v-\alpha)(R-\beta)\}.
\label{eq:Bteps}
\end{equation}
The bound is attained by the three-point measure with masses
\begin{equation}
p_R=p^{\ast},\qquad
p_\beta=\frac{v-\alpha-p^{\ast}(R-\alpha)}{\beta-\alpha},\qquad
p_\alpha=1-p_\beta-p_R
\label{eq:Bthreepoint}
\end{equation}
on $R$, $\beta$, and $\alpha$, where
$p^{\ast}=\min\{\epsilon,(v-\alpha)/(R-\alpha)\}$ is the largest tail weight compatible with
$y\ge\alpha$ once the tail sits at $R$.  Both remaining masses are nonnegative:
$p_\beta\ge0$ because $p^{\ast}\le(v-\alpha)/(R-\alpha)$, and, after eliminating $p_R$,
\begin{equation}
p_\alpha=\frac{\beta-v+p^{\ast}(R-\beta)}{\beta-\alpha}\ \ge\ 0
\label{eq:Bpalpha}
\end{equation}
because $v\le\beta\le R$.  The second branch of Eq.~(\ref{eq:Bteps}) corresponds to
$p_\beta=0$, the two-point measure on $\{\alpha,R\}$.  As $\epsilon\to0$ the tail weight
vanishes and $U_\epsilon\to U_{\mathrm{BD}}$.

\subsection{Mean and second moment}  For
\begin{equation}\mathcal C_{m_2}=\{\mu:\int y\,d\mu=v,\ \int y^{2}d\mu=m_2\}\end{equation} the objective takes the single
value $m_2$ on the whole class, so
\begin{equation}
U^{\ast}=\frac{m_2}{D}=\dot\sigma
\label{eq:Bm2}
\end{equation}
exactly.  The class itself is not a singleton. With the mean and the second moment fixed, and
provided $m_2\ge v^{2}$ and the pair $(v,m_2)$ is compatible with the support, infinitely many
distributions are admissible.  

\subsection{Dynamics-constrained class}  A class $\mathcal C_{\mathrm{dyn}}$ would further
restrict $\mu$ to distributions realized by the stationary Fokker--Planck dynamics.  For a
one-signed current with $p=(1+c)/2\pi$ the steady state satisfies the exact identity
$\langle1/y\rangle=(1+M_2)/v$ with $M_2=\langle c^{2}\rangle$, which is not a moment
constraint, and the corresponding reachability problem is left to future work.

\section{The affine family and endpoint optimality}\label{app:affine}
For $f(y)=y^{2}$ a bound proved on $[0,1]$ carries
over to $[\alpha,\beta]$ through $y=\alpha+(\beta-\alpha)y'$, which maps the mean $v$ to
$v'=(v-\alpha)/(\beta-\alpha)$ and rescales the second moment, so it suffices to work with
$\alpha=0$, $\beta=1$, where $f(0)=0$ and $f(1)=1$.

Let $a=1-b-c$ and let the parameters $(b,c)$ lie in the set
\begin{equation}
0\le b\le 1-c\quad(0\le c\le1),\qquad
-c\le b\le1\quad(-1\le c<0),
\label{eq:Crange}
\end{equation}
which keeps $a\ge0$ in both branches and reduces to $b=0$ at $c=1$ and to $b=1$ at $c=-1$.  Note
that $a+b+c=1$ holds identically, so the coefficients always form a partition of
unity. 
That $a\ge0$ by construction and $b\ge\max\{0,-c\}\ge0$ in both branches, so only the sign
of $c$ changes, continuously, at $c=0$. 

On
this set the affine Jensen--Mercer inequality holds:
\begin{equation}
(b+cy)^{2}\le b+c\,y^{2},\qquad y\in[0,1]
\label{eq:Cpoint}
\end{equation}
For $0\le c\le1$ it is the weighted Jensen inequality \cite{Jensen06,Pecaric}
\begin{equation}f(a\cdot0+b\cdot1+cy)\le af(0)+bf(1)+cf(y)\end{equation} evaluated at the affine argument $b+cy$ with weights
$(a,b,c)$. The branch $-1\le c<0$ is its reflection extension, and the endpoint $b=1$, $c=-1$ is
Mercer's inequality \cite{Mercer03}, 
\begin{equation}(1-y)^{2}\le1-y^{2}\end{equation} with equality at both endpoints of
$[0,1]$.  

 Averaging Eq.~(\ref{eq:Cpoint}) over the distribution of the local
current and using Jensen's inequality on the left,
\begin{equation}
(b+cv)^{2}\le\big\langle(b+cy)^{2}\big\rangle\le b+c\,\langle y^{2}\rangle ,
\label{eq:Cavg}
\end{equation}
so that
\begin{equation}
c\,\langle y^{2}\rangle\ \ge\ (b+cv)^{2}-b .
\label{eq:Cbound}
\end{equation}
For $c>0$ this is a lower bound on the dissipation, and for $c<0$ the upper bound
\begin{equation}
\langle y^{2}\rangle\ \le\ U(b,c)\equiv\frac{(b+cv)^{2}-b}{c},\qquad c<0 .
\label{eq:CU}
\end{equation}

 For fixed $c$ one has \begin{equation}U=g(b)/c,\quad\text{with},\quad g(b)=(b+cv)^{2}-b\end{equation} convex, so
the minimum of $U$ over $b\in[-c,1]$ is attained at an endpoint,
\begin{equation}
U(-c,c)=1+c(1-v)^{2},\qquad U(1,c)=2v+cv^{2}.
\label{eq:Cend}
\end{equation}
Both values increase with $c$ and they coincide at $c=-1$, where the admissible set shrinks
to the single point $b=1$.  Hence the sharpest upper bound in the family is
\begin{equation}
U_{\mathrm{aff}}(v)=2v-v^{2}\qquad(c=-1,\ b=1).
\label{eq:Cmercer}
\end{equation}

 For $c>0$ the bound is maximized at an endpoint of $b\in[0,1-c]$,
\begin{equation}
U(0,c)=cv^{2},\qquad U(1-c,c)=2v-1+c(1-v)^{2},
\label{eq:Clow}
\end{equation}
and since $cv^{2}\le v^{2}$ and $2v-1+c(1-v)^{2}\le v^{2}$ the sharpest lower bound in the
family is $v^{2}$, attained at $c=1$ and $b=0$.  It is the Jensen floor.

The family thus brackets the dissipation as
\begin{equation}
v^{2}\ \le\ \dot\sigma\ \le\ 2v-v^{2},
\label{eq:Cfam}
\end{equation}
with the lower endpoint at $c=1$ and the upper endpoint at $c=-1$, and every interior member
is weaker on at least one side.  At $v=\tfrac12$, for instance, the member $c=-\tfrac12$
gives $U=\tfrac78$, against $U=\tfrac34$ at $c=-1$.  The upper endpoint is itself the
non-sharp bound obtained by estimating the reflected second moment with Jensen's inequality,
\begin{equation}
2v-v^{2}-v=v(1-v)=\mathrm{Var}_{\max},
\label{eq:Cgap}
\end{equation}
so it lies above the sharp ceiling of Eq.~(\ref{eq:upper}) by exactly
$\mathrm{Var}_{\max}$.  The affine family therefore reproduces the Jensen floor and the
non-sharp Mercer-type ceiling, and no member improves on the two-sided sandwich.

\section{Dynamical attainability and the bounded-drive ceiling in one dimension}\label{app:dyn}

\subsection{Riccati reduction}  
On the ring $[0,2\pi)$ with constant $D$ a steady state has a
constant flux $J$.  Writing the local current as $y=J/p$,  the stationary Fokker--Planck equation $J=\mu fp-Dp'$ becomes
\begin{equation}1=\mu f/y+Dy'/y^{2},\end{equation} which is Eq.~(\ref{eq:riccati}).
Conversely, any positive $C^{1}$, $2\pi$-periodic $y$ is produced by the drift 
\begin{equation}
f=\frac{1}{\mu}\Big(y-D\frac{y'}{y}\Big),\qquad J=\Big(\int_0^{2\pi}\frac{dx}{y}\Big)^{-1}.
\label{eq:inverse}
\end{equation}
  Every smooth drive has the conservative form $f=F-V'$ with a periodic $V$.  Integrating
\eqref{eq:riccati} over the ring gives the normalization $(2\pi)^{-1}\int dx/y=1/v$ and, since
$\int dx\,y'/y=0$ for a periodic current, the arithmetic mean \eqref{eq:arithmean}.  With
$\dot\sigma=D^{-1}\langle y^{2}\rangle=(v/2\pi D)\int_0^{2\pi}y\,dx$ these combine into
\eqref{eq:EPflux}.

\subsection{The supremum is unchanged}  Fix $0<\alpha\le v\le\beta$ and let $y$ be positive,
$2\pi$-periodic, with values in $[\alpha,\beta]$ and harmonic mean
$(2\pi)^{-1}\int dx/y=1/v$.  The pointwise inequality
\begin{equation}
(\alpha+\beta-y)-\frac{\alpha\beta}{y}=-\frac{(y-\alpha)(y-\beta)}{y}\ \ge\ 0
\label{eq:Achordharm}
\end{equation}
gives $y\le\alpha+\beta-\alpha\beta/y$.  Averaging over the ring gives
\begin{equation}
\langle y\rangle_{\mathrm{sp}}\le\alpha+\beta-\frac{\alpha\beta}{v},
\qquad
\dot\sigma\le\frac{v}{D}\Big(\alpha+\beta-\frac{\alpha\beta}{v}\Big)
=U_{\mathrm{BD}},
\label{eq:Aharmbound}
\end{equation}
and equality holds only when $y$ takes the two values $\alpha$ and $\beta$.
The two-level profile saturates this bound.  Put $y=\beta$ on a fraction $f_\beta$ of the ring and
$y=\alpha$ elsewhere.  Its harmonic mean is $f_\beta/\beta+(1-f_\beta)/\alpha=1/v$ at
$f_\beta=\beta(v-\alpha)/[v(\beta-\alpha)]$, and its arithmetic mean is then
$\alpha+f_\beta(\beta-\alpha)=\alpha+\beta-\alpha\beta/v$, which gives $\dot\sigma=U_{\mathrm{BD}}$
through \eqref{eq:Aharmbound}.

This profile is not $C^{1}$, so we smooth it.  For $\delta>0$ let $\psi_{f,\delta}$ equal $1$ on an
arc of length $2\pi f-\delta$, equal $0$ on an arc of length $2\pi(1-f)-\delta$, and join the two
arcs monotonically across two transitions of width $\delta$, symmetric about their midpoints as
linear ramps are, so that the three lengths add up to the circumference.  The profile exists for
$0<\delta<\pi\min\{f,1-f\}$, and nested arcs make $\psi_{f,\delta}\le\psi_{f',\delta}$ pointwise
for $f\le f'$.  Set $y_{f,\delta}=\alpha+(\beta-\alpha)\psi_{f,\delta}$.  Then
\begin{equation}
H(f)\equiv(2\pi)^{-1}\int_0^{2\pi}\frac{dx}{y_{f,\delta}}
=\frac{f}{\beta}+\frac{1-f}{\alpha}+O(\delta),
\label{eq:Hf}
\end{equation}
uniformly in $f$ and $\delta$.  At fixed $\delta$ the map $f\mapsto H(f,\delta)$ decreases
strictly, by the nesting property.  The interval $[H(1,\delta),H(0,\delta)]$ contains $1/v$ once
$3\delta/(2\pi\alpha)<\min\{1/\alpha-1/v,1/v-1/\beta\}$, so a unique $f(\delta)$ satisfies
$H(f(\delta),\delta)=1/v$.  Comparing with $\varphi(f_\beta)=1/v$, where
$\varphi(f)=f/\beta+(1-f)/\alpha$, gives
\begin{equation}
|f(\delta)-f_\beta|\ \le\ \frac{3\delta}{2\pi\alpha}\cdot\frac{\alpha\beta}{w}
=C\delta,\qquad C\equiv\frac{3\beta}{2\pi(\beta-\alpha)},
\label{eq:Cest}
\end{equation}
and $C$ does not depend on $\delta$, on $f$ or on the shape of the transitions.  Take $\delta$ so
small that $C\delta<\tfrac12m_\beta$ and $\delta<\tfrac12\pi m_\beta$, with
$m_\beta=\min\{f_\beta,1-f_\beta\}$.  Then both plateaus stay positive and every step above
remains valid.  A symmetric transition contributes exactly half its width to the arithmetic mean,
so the two plateau reductions and the two transitions cancel and
$\langle y_{f(\delta),\delta}\rangle=\alpha+(\beta-\alpha)f(\delta)$.  This tends to
$\alpha+\beta-\alpha\beta/v$, hence $\dot\sigma\to U_{\mathrm{BD}}$ along a family of $C^{1}$
currents.

\subsection{Fixed tilt}  If the tilt $\bar f$ is known, Eq.~(\ref{eq:arithmean}) fixes the arithmetic
mean as well, $\langle y\rangle_{\mathrm{sp}}=\mu\bar f$, and Eq.~(\ref{eq:EPflux}) fixes
$\dot\sigma$, so the tightness ratio collapses to Eq.~(\ref{eq:rhoTilt}) and the data
$\{v,\alpha,\beta\}$ are LP-tight exactly when $\mu\bar f=(\alpha+\beta)-\alpha\beta/v$.  The
two-level family that carries both means is parametrized by the fraction $t$ of the ring at the
lower level, with $\alpha$ the root in $(0,v)$ of
\begin{equation}
t\,\alpha^{2}+\big[v(1-2t)-\mu F\big]\alpha+t\,\mu F\,v=0,
\label{eq:twolevel}
\end{equation}
$\beta=(\mu F-t\alpha)/(1-t)>v$, the discriminant $(\mu F-v)^{2}+4t(1-t)v(\mu F-v)$ positive, and
every member tight.

The rounding above leaves the arithmetic mean at $\mu F+a_\delta$ with
$|a_\delta|\le(\beta-\alpha)C\delta$, since $f(\delta)=f_\beta+O(\delta)$, while the harmonic mean
is $1/v$ by construction.  Perturb $y=y_{f(\delta),\delta}+a+b\xi$, with $\xi$ a fixed smooth
$2\pi$-periodic function such that $\int\xi\,dx=0$ and $\avg{\xi/y_{f,\delta}^{2}}\ne0$.  The map
$(a,b)\mapsto(\avg{y}-\mu F,\avg{1/y}-1/v)$ vanishes at the origin up to $O(\delta)$ and has there
the Jacobian
\begin{equation}
\begin{pmatrix}1&0\\ -\avg{y_{f,\delta}^{-2}}&-\avg{\xi\,y_{f,\delta}^{-2}}\end{pmatrix},
\end{equation}
whose determinant $-\avg{\xi/y_{f,\delta}^{2}}$ is nonzero.  The implicit function theorem gives a
unique $(a,b)=O(\delta)$ with both means exact.  The perturbed profile is positive and $C^{1}$,
because $y_{f,\delta}\ge\alpha$ and $|a|,|b|=O(\delta)$, its range is $[\alpha,\beta]$ up to
$O(\delta)$, and $\avg{y^{2}}\to\avg{y_0^{2}}$, so its tightness ratio tends to one.

\subsection{Slope majorant}  Since \eqref{eq:riccati} reads $Dy'=y^{2}-\mu fy$, we have
\begin{equation}
D\,|y'|\ \le\ \kappa\equiv\max_{0\le x<2\pi}\big|y^{2}-\mu fy\big| ,
\label{eq:slopebound}
\end{equation}
and a bound on the drive alone converts this into an explicit majorant.  For a drive interval
$f_{-}\le f\le f_{+}$,
\begin{equation}
\begin{split}
\kappa\ \le\ \bar\kappa\equiv\max_{f\in\{f_{-},f_{+}\}}\max\big\{&
\alpha\big|\alpha-\mu f\big|,\ \beta\big|\beta-\mu f\big|,\\
&\tfrac14(\mu f)^{2}\ \ \text{if}\ \ \alpha\le\tfrac12\mu f\le\beta\big\}.
\end{split}
\label{eq:kapbar}
\end{equation}
For a symmetric constraint
$|f|\le F_{\max}$ the maximum is known in closed form.  Since $y>0$,
\begin{equation}
|y^{2}-\mu fy|\ \le\ y^{2}+\mu F_{\max}y\ \le\ \beta^{2}+\mu F_{\max}\beta
\qquad\big(y\in[\alpha,\beta]\big),
\label{eq:kapmax}
\end{equation}
with equality at $y=\beta$, $f=-F_{\max}$, so that $\bar\kappa=\beta(\beta+\mu F_{\max})$
exactly for every admissible range and every $F_{\max}\ge0$.  

\subsection{Weighted gap and level integral}  With $\alpha=\min y$ and $\beta=\max y$ attained on the
ring, and $y^{2}-(\alpha+\beta)y+\alpha\beta=-(y-\alpha)(y-\beta)$, we have the average
\begin{equation}
\big\langle (y-\alpha)(\beta-y)\big\rangle
=J\int_{0}^{2\pi}\frac{(y-\alpha)(\beta-y)}{y}\,dx ,
\label{eq:weightedgap}
\end{equation}
whose integrand $g(y)=(y-\alpha)(\beta-y)/y$ is positive on $(\alpha,\beta)$.  Its level integral
is
\begin{equation}
\mathcal J(\alpha,\beta)\equiv\int_{\alpha}^{\beta}g\,dy
=\frac{\beta^{2}-\alpha^{2}}{2}-\alpha\beta\ln\frac{\beta}{\alpha},
\label{eq:Jdef}
\end{equation}
and on $(\alpha,\beta)$ one has $g\ge(y-\alpha)(\beta-y)/\beta$, so that
\begin{equation}
\mathcal J(\alpha,\beta)\ \ge\ \frac{w^{3}}{6\beta},\qquad w\equiv\beta-\alpha\le\beta ,
\label{eq:Jband}
\end{equation}
the first inequality being asymptotically sharp as $w/\beta\to0$.

\subsection{The weighted crossing bound}  Let $y\in C^{1}$ be positive and $2\pi$-periodic, with
$\alpha=\min y$ and $\beta=\max y$ both attained, $K=\max|y'|$, and
$g(y)=(y-\alpha)(\beta-y)/y\ge0$ on $[\alpha,\beta]$.  If $K=0$ the current is constant and both
sides of \eqref{eq:boundeddrive} equal $v^{2}/D$, so assume $K>0$.  Since $\alpha>0$ the factor
$1/y$ is bounded and $\mathcal J$ is finite.  Cover the ring by the maximal intervals $I_k$ on
which $y$ is strictly monotone.  Substituting $y$ for $x$,
\begin{equation}
\int_{I_k}g(y(x))\,dx=\int_{y(I_k)}\frac{g(y)}{|y'(y)|}\,dy\ \ge\ \frac1K\int_{y(I_k)}g(y)\,dy ,
\label{eq:appcross1}
\end{equation}
and the sum over $k$ is organized through the area formula
\begin{equation}
\int_0^{2\pi}g(y)\,|y'|\,dx=\int_\alpha^\beta g(t)\,N(t)\,dt ,
\label{eq:appcross3}
\end{equation}
where $N(t)$ is the number of solutions of $y(x)=t$. Combined with $|y'|\le K$ and $N\ge2$ this gives
\begin{align}
\int_0^{2\pi}g(y(x))\,dx\ &\ge\ \frac1K\int_0^{2\pi}g(y)\,|y'|\,dx\nonumber\\
&=\frac1K\int_\alpha^\beta g(t)\,N(t)\,dt\nonumber\\
&\ge\ \frac{2}{K}\int_\alpha^\beta g(t)\,dt\ =\ \frac{2\,\mathcal J}{K}.
\label{eq:appcross2}
\end{align}
 Finally, the slope bound \eqref{eq:slopebound} gives
$K\le\bar\kappa/D$, and inserting this with $J=v/2\pi$ into Eq.~(\ref{eq:weightedgap}) yields
Eq.~(\ref{eq:boundeddrive}).

\section{Underdamped dynamics: Cross terms, moment hierarchy, and Hermite expansion}\label{app:under}\label{app:moments}

\subsection{Cross terms}  With $\nu_i(x)=\int v_i\,p(v_i|x)\,dv_i$, write
$u_i=\nu_i+[v_i-\nu_i+\tfrac{k_BT_i}{m_i}\partial_{v_i}\ln p]$.  Then
$\avg{\nu_i(v_i-\nu_i)}=0$ by the definition of $\nu_i$, and
$\tfrac{k_BT_i}{m_i}\avg{\nu_i\,\partial_{v_i}\ln p}=0$ is a total $v$ derivative, so
\begin{equation}\avg{u_i^{2}}=\avg{\nu_i^{2}}+\avg{[v_i-\nu_i+\tfrac{k_BT_i}{m_i}\partial_{v_i}\ln p]^{2}},\end{equation} which is Eq.~(\ref{eq:exactdecomp}).
Moreover $\avg{u_i}=\avg{v_i}$, since the boundary term vanishes.  Writing
$p=e^{-mv^{2}/2k_BT}\Psi$ gives $u=\tfrac{k_BT}{m}\partial_v\ln\Psi$, bounded if and only if
$\partial_v\ln\Psi$ is.

\subsection{Moment hierarchy}  Multiplying Eq.~(\ref{eq:kramers}) by $v^{k}$ for a single
coordinate and integrating gives
\begin{equation}
0=-\partial_x m_{k+1}+\frac{k}{m}(f m_{k-1}-\zeta m_k)
+\frac{k_BT\zeta}{m^{2}}k(k-1)m_{k-2},
\label{eq:Amomhier}
\end{equation}
which for $k=0,\dots,4$ reads
\begin{equation}
\begin{aligned}
m_1&=J,\\
m_1&=\zeta^{-1}(fm_0-m\partial_xm_2),\\
m_2&=\theta m_0+\zeta^{-1}fm_1-\tfrac{m}{2\zeta}\partial_xm_3,\\
m_3&=\zeta^{-1}fm_2+2\theta m_1-\tfrac{m}{3\zeta}\partial_xm_4,\\
m_4&=3\theta m_2+O(\zeta^{-2}).
\end{aligned}
\label{eq:Amom45}
\end{equation}
Substituting $m_2=\theta m_0+O(\zeta^{-2})$ and
$m_4=3\theta m_2+O(\zeta^{-2})=3\theta^{2}m_0+O(\zeta^{-2})$ into the $k=3$ relation gives
\begin{equation}
m_3=\frac{\theta}{\zeta}\big(fm_0-k_BT\partial_xm_0\big)+2\theta m_1+O(\zeta^{-2}),
\end{equation}
and since $fm_0-k_BT\partial_xm_0=\zeta m_1+O(\zeta^{-2})$ while $m_1=J$ is $x$ independent in
a steady state, $m_3=3\theta J+O(\zeta^{-2})$ and hence $\partial_xm_3=O(\zeta^{-2})$.
Therefore
\begin{equation}
\begin{split}
S-\theta&=\frac{m_2}{m_0}-\theta-\nu^{2}
=\frac{f\nu}{\zeta}-\nu^{2}-\frac{m}{2\zeta}\frac{\partial_xm_3}{m_0}\\
&=\frac{f\nu}{\zeta}-\nu^{2}+O(\zeta^{-3}),
\end{split}
\label{eq:ASdev}
\end{equation}
which with the Smoluchowski relation
$\nu=\zeta^{-1}(f-k_BT\partial_x\ln n)+O(\zeta^{-3})$ reads Eq.~(\ref{eq:Sdev}).  

\subsection{Hermite expansion and the adiabatic correction}  Expanding the velocity distribution
around the local Maxwellian, $p=n\phi(1+\psi)$ with
$\phi=(2\pi\theta)^{-1/2}e^{-v^{2}/2\theta}$ and $\theta=k_BT/m$, the irreversible current takes
the form $u=\theta\,\partial_v\ln(1+\psi)$.  Writing
$\psi=\psi_1+\psi_2+O(\zeta^{-3})$ in Hermite polynomials $\mathrm{He}_j(v/\sqrt\theta)$
orthogonal with respect to $\phi$, the leading coefficient is
$c_1=\sqrt{mk_BT}\,\zeta^{-1}(\beta f-\partial_x\ln n)$, so that
$\theta\,\partial_v\psi_1=(k_BT/\zeta)(\beta f-\partial_x\ln n)\equiv\nu^{(1)}$ equals the
Smoluchowski velocity and
\begin{equation}
u=\nu^{(1)}+\theta\,\partial_v\psi_2-\theta\,\psi_1\partial_v\psi_1+O(\zeta^{-3}).
\label{eq:Auherm}
\end{equation}
The term linear in $\psi_1$ integrates to zero,
$\avg{\nu^{(1)}\psi_1\partial_v\psi_1}=0$, and the remaining $O(\zeta^{-3})$ cross term
cancels against the renormalization of the mean current
$\nu=\avg{u}=\nu^{(1)}+c_{12}\sqrt\theta+O(\zeta^{-3})$, where $c_{12}$ is the $\mathrm{He}_1$
coefficient of $\psi_2$.  The result is
\begin{equation}
\avg{u^{2}}=\avg{\nu^{2}}+\theta^{2}\avg{(\partial_v\psi_2)^{2}}+O(\zeta^{-4}),
\label{eq:Ausq}
\end{equation}
with
$\theta^{2}\avg{(\partial_v\psi_2)^{2}}=\theta\avg{c_{12}^{2}}+4\theta\avg{c_2^{2}}
+18\theta\avg{c_3^{2}}+\cdots\ge0$, which vanishes only for shifted Maxwellians.  Since
$\partial_v\psi_2=O(\zeta^{-2})$ and $\nu=O(\zeta^{-1})$, the relative correction is
$O(\varepsilon^{2})$ with $\varepsilon^{2}=mk_BT/(\zeta^{2}L^{2})=\tau_p/\tau_D$, which is
Eq.~(\ref{eq:adiabatic}).

 The order statement
$\dot\sigma_u=\dot\sigma_{\mathrm{ov}}[1+O(\varepsilon^{2})]$ is a conditional adiabatic
theorem.  It holds when (i) the potential force $f$ and the spatial derivatives of the steady
density are bounded uniformly in $\zeta$ as $\zeta\to\infty$ with $m$, $k_BT$, the length
scale $L$, and $f$ held fixed, when (ii) the boundary conditions, for instance periodic ones,
make the steady state solvable with $\partial_xJ=0$ and the Smoluchowski equation solvable,
and when (iii) $\mu_1^{(0)}=fm_0-k_BT\partial_xm_0=\zeta J+O(\zeta^{-2})$ with $J$ constant,
which forces $m_3=3\theta J+O(\zeta^{-2})$ and $\partial_xm_3=O(\zeta^{-2})$.  These conditions
control the remainder $O(\zeta^{-2})$ of the velocity-variance deficit
\eqref{eq:Sdev} and the remainder $O(\zeta^{-4})$ of the excess dissipation
\eqref{eq:Ausq}. The relative error of the rate is $O(\varepsilon^{2})$ throughout.

\section{Tilted periodic potential: Full computation}\label{app:tilted}

\subsection{Fourier solution}  With $V=V_0\cos x$ and $J=D\beta(F-V')p-Dp'$, Fourier expanding
$p=(2\pi)^{-1}(1+\sum_{n\ne0}\hat p_ne^{inx})$ with $\hat p_0=1$ gives, for $n\ne0$,
\begin{equation}
(a-in)\hat p_n+\frac{g}{2i}(\hat p_{n-1}-\hat p_{n+1})=0,
\end{equation}
together with the flux condition
\begin{equation}
2\pi J=Da+D\frac{g}{2i}(\hat p_{-1}-\hat p_1),
\end{equation}
with $a=\beta F$ and $g=\beta V_0$.  Solving order by order,
\begin{equation}
\begin{aligned}
u_1&=\frac{-1}{2i(a-i)}=\frac{-1+ia}{2(1+a^{2})},\\
v_2&=\frac{-u_1}{2i(a-2i)}
=\frac{2-a^{2}-3ia}{4(1+a^{2})(a^{2}+4)},\\
u_3&=\frac{v_2}{2i(a-i)}
=\frac{(2-4a^{2})+i(a^{3}-5a)}{8(1+a^{2})^{2}(a^{2}+4)},\\
w_3&=\frac{-v_2}{2i(a-3i)}
=\frac{6(a^{2}-1)+ia(11-a^{2})}{8(1+a^{2})(a^{2}+4)(a^{2}+9)},\\
v_4&=\frac{-(u_3-w_3)}{2i(a-2i)}
=\frac{(-a^{4}+17a^{2}-6)+i(17a-7a^{3})}{8(1+a^{2})^{2}(a^{2}+4)(a^{2}+9)},\\
x_4&=\frac{-w_3}{2i(a-4i)}
=\frac{(a^{4}-35a^{2}+24)+i(10a^{3}-50a)}{16(1+a^{2})(a^{2}+4)(a^{2}+9)(a^{2}+16)}.
\end{aligned}
\end{equation}
Since $\hat p_{-1}-\hat p_1=-2i\,\mathrm{Im}\,\hat p_1$ and
\begin{equation}
\mathrm{Im}\,\hat p_1
=g\frac{a}{2(1+a^{2})}
+g^{3}\frac{a^{3}-5a}{8(1+a^{2})^{2}(a^{2}+4)}+O(g^{5}),
\end{equation}
the current is
\begin{equation}
J=\frac{D}{2\pi}\Big[a-\frac{a\,g^{2}}{2(1+a^{2})}
+\frac{a(5-a^{2})g^{4}}{8(1+a^{2})^{2}(a^{2}+4)}+O(g^{6})\Big].
\label{eq:AJ}
\end{equation}

\subsection{Moments and the exact rate}  For $M_1$, with the angle brackets here denoting the
uniform average $(1/2\pi)\int_0^{2\pi}dx$ over the ring,
\begin{equation}
\begin{aligned}
M_1&=1+\avg{c^{2}}-\avg{c^{3}}+\avg{c^{4}}+O(g^{6}),
\\
\avg{c^{2}}&=\sum_{n\ne0}|\hat p_n|^{2}
=\frac{g^{2}}{2(1+a^{2})}\\
&\quad+\frac{3(a^{2}-1)g^{4}}{8(1+a^{2})^{2}(a^{2}+4)}+O(g^{6}),\\
\avg{c^{3}}&=3\hat p_1^{2}\hat p_{-2}+3\hat p_2\hat p_{-1}^{2}+O(g^{6})
=6g^{4}\,\mathrm{Re}\!\left(u_1^{2}\overline{v_2}\right)+O(g^{6})\\
&=\frac{3(a^{2}+2)g^{4}}{8(1+a^{2})^{2}(a^{2}+4)}+O(g^{6}),\\
\avg{c^{4}}&=6|\hat p_1|^{4}+O(g^{6})\\
&=\frac{3g^{4}}{8(1+a^{2})^{2}}+O(g^{6}),
\end{aligned}
\end{equation}
which combine to
\begin{equation}
M_1=1+\frac{g^{2}}{2(1+a^{2})}+\frac{3\,g^{4}}{8(1+a^{2})(a^{2}+4)}+O(g^{6});
\label{eq:AM1}
\end{equation}
the $\avg{c^{3}}$ contribution is essential.  Finally,
\begin{equation}
\dot\sigma=D^{-1}J^{2}\!\int p^{-1}dx=D^{-1}J^{2}\,4\pi^{2}M_1 .
\end{equation}
\subsection{Thermodynamic cross-check}  Since the tilt performs the only net work,
$\dot\sigma=Fv/(k_BT)$ with $v=2\pi J$.  Equating this with the expression above yields the
exact relations
\begin{equation}
J=\frac{Da}{2\pi M_1},\qquad \dot\sigma=\frac{Da^{2}}{M_1},
\label{eq:Aexact}
\end{equation}
i.e., Eq.~(\ref{eq:tiltedexact}).  Expanding $M_1^{-1}$ gives Eq.~(\ref{eq:sigexp}).

\subsection{Strong-barrier asymptotics}  For $g\gg1$ at fixed $a>0$ the current is
\begin{equation}J/D\approx(g/2\pi)(e^{a\pi}-e^{-a\pi})e^{-2g},\end{equation} and the steady solution can be written as 
\begin{equation}
p(x)=e^{ax-g\cos x}\,\frac{J}{D}\,\big[C_0-\mathrm{cf}(x)\big],
\label{eq:strongp}
\end{equation}
with $C_0=I_0\,e^{2\pi a}/(e^{2\pi a}-1)$, $I_0=\int_0^{2\pi}e^{-as+g\cos s}ds$, and
$\mathrm{cf}(x)=\int_0^{x}e^{-as+g\cos s}ds$.  Both integrals are dominated by the saddle at
$s=0$, and Laplace's method gives, away from the barrier top $x=0$ and from the minimum of
$p$,
\begin{equation}
p(x)\approx\frac{J}{D}\,\sqrt{\frac{2\pi}{g}}\,
\frac{e^{g(1-\cos x)+ax}}{e^{2\pi a}-1}.
\label{eq:Astrongp}
\end{equation}
At the well bottom $x=\pi$ this gives $p(\pi)\simeq\sqrt{g/(2\pi)}$, hence
\begin{equation}
u_{\min}=\frac{1}{p(\pi)}=\sqrt{\frac{2\pi}{g}}\,\big[1+O(1/g)\big].
\label{eq:Aumin}
\end{equation}

The maximum of $u$ is not at the barrier top.  For $x=O(g^{-1/2})$ the upper limit of
$\mathrm{cf}(x)$ lies inside the Gaussian width of the saddle, so the integral is cut by its
lower limit and an error function survives.  With $x=\xi\sqrt{2/g}$,
\begin{equation}
p\simeq\frac{J}{D}\sqrt{\frac{\pi}{2g}}\,
e^{\xi^{2}}\big[\coth(\pi a)-\mathrm{erf}(\xi)\big].
\label{eq:Aboundary}
\end{equation}
 Its positive root $\xi^{*}(a)$
locates the minimum of $p$ at $x^{*}=\xi^{*}(a)\sqrt{2/g}+O(1/g)$, and evaluated there
Eq.~(\ref{eq:Aboundary}) gives Eq.~(\ref{eq:umax}) with $\gamma_1(a)=1/[2\xi^{*}(a)]$.

With $y=J/p\in[Ju_{\min},Ju_{\max}]$ and $\langle y\rangle=2\pi J$, the ceiling of
Eq.~(\ref{eq:upper}) reads
\begin{equation}
U_{\mathrm{BD}}=J^{2}\big[2\pi(u_{\min}+u_{\max})-u_{\min}u_{\max}\big],
\label{eq:AUbd}
\end{equation}
and Eq.~(\ref{eq:Aexact}) gives $\dot\sigma=Da^{2}/M_1$, so that
\begin{equation}
\begin{split}
\frac{U_{\mathrm{BD}}}{\dot\sigma}
&=\frac{2\pi(u_{\min}+u_{\max})-u_{\min}u_{\max}}{4\pi^{2}M_1}\\
&=\frac{4\pi^{2}+(2\pi-u_{\min})(u_{\max}-2\pi)}{4\pi^{2}M_1},
\end{split}
\label{eq:Agap}
\end{equation}
which is Eq.~(\ref{eq:stronggap}).  With
$u_{\max}/(2\pi M_1)=\sqrt{g/2}/(a\gamma_1)$ and
$M_1\approx a\,e^{2g+a\pi}/[g(e^{2\pi a}-1)]$, Eq.~(\ref{eq:Agap}) reduces at leading order
to Eq.~(\ref{eq:stronggap}), the additive constant being set by the $u_{\min}$ correction.
The tightness ratio follows from
\begin{equation}\rho=\frac{4\pi^{2}(M_1-1)}{(u_{\max}-2\pi)(2\pi-u_{\min})}\simeq2\pi M_1/u_{\max},\end{equation} which is
Eq.~(\ref{eq:strongrho}).

\subsection{Tightness ratio}  With $u=1/p=2\pi(1+c)^{-1}$, the exact identities
$\avg{u}=\int(1/p)\,p\,dx=2\pi$ and
$\avg{u^{2}}=\int(1/p^{2})\,p\,dx=\int(1/p)\,dx=4\pi^{2}M_1$ give
\begin{multline}
\mathrm{Var}(u)=4\pi^{2}(M_1-1)\\
=(2\pi)^{2}\Big[\frac{g^{2}}{2(1+a^{2})}
+\frac{3\,g^{4}}{8(1+a^{2})(a^{2}+4)}+O(g^{6})\Big].
\end{multline}
For the denominator, write 
\begin{equation}c=gA+g^{2}B+g^{3}C+g^{4}D+O(g^{5})\end{equation} with
$A=2\,\mathrm{Re}\,u_1e^{ix}=-\lambda\cos(x-\varphi)$,
$\lambda=(1+a^{2})^{-1/2}$, $\tan\varphi=a$,
$B=2\,\mathrm{Re}\,v_2e^{2ix}$,
$C=2\,\mathrm{Re}(u_3e^{ix}+w_3e^{3ix})$, and
$D=2\,\mathrm{Re}(v_4e^{2ix}+x_4e^{4ix})$, so that
\begin{equation}u-2\pi=2\pi[-gA+g^{2}(A^{2}-B)+g^{3}(2AB-A^{3}-C)+O(g^{4})].\end{equation}
The maximum of $u$ sits at $x_{+}=\varphi+\delta_1g+O(g^{2})$, where $A=-\lambda$, and
the minimum at $x_{-}=\varphi+\pi+\tilde\delta_1g+O(g^{2})$, where $A=+\lambda$.  The
condition $c'(x)=0$ gives $\delta_1=-\tilde\delta_1=-B'_0/\lambda$ with
$B'_0=B'(\varphi)=-a/[(1+a^{2})(a^{2}+4)]$.  Expanding each one-sided deviation to
$O(g^{4})$ and collecting the product
$\mathrm{Var}_{\max}=(2\pi)^{2}[\lambda g+U_2g^{2}+U_3g^{3}+\cdots][\lambda g+V_2g^{2}+V_3g^{3}
+\cdots]$ gives, in terms of the values at the two extrema,
\begin{equation}
U_2=V_2=\lambda^{2}-B_0,\qquad
\delta_1=-\frac{B'_0}{\lambda},\qquad
\tilde\delta_1=-\delta_1 ,
\end{equation}
with $B_0=B(\varphi)$ and $B'_0=B'(\varphi)$. The equality $U_2=V_2$ is exactly the statement
that the odd coefficients cancel, so that
\begin{equation}
\begin{split}
&\mathrm{Var}_{\max}=(2\pi)^{2}\big[\lambda^{2}g^{2}+q_4\,g^{4}+O(g^{6})\big],\\
&q_4=\lambda(U_3-V_3)-(\lambda^{2}-B_0)^{2},
\end{split}
\end{equation}
and the remaining collection uses only the two explicit values
\begin{equation}
B_0=\frac{a^{2}+2}{2(1+a^{2})(a^{2}+4)},\qquad
B'_0=-\frac{a}{(1+a^{2})(a^{2}+4)},
\end{equation}
together with the third-harmonic shift $C_0-C_0^{-}$.  This yields
\begin{equation}
q_4=\frac{3s^{3}+35s^{2}+156s+156}{4(1+s)^{2}(s+4)^{2}(s+9)},
\qquad s=a^{2}.
\label{eq:q4}
\end{equation}
  Carrying the same recursion one order further gives the
$g^{6}$ coefficients in closed form.  With $s=a^{2}$,
\begin{equation}
\begin{split}
m_6&=\frac{5}{16(1+s)(s+4)(s+9)},\\
q_6&=\big(5s^{8}+276s^{7}+6764s^{6}+85526s^{5}+609567s^{4}\\
&\qquad+2588102s^{3}+5671680s^{2}+6003360s+2085120\big)\\
&\qquad\big/\big[8(1+s)^{3}(s+4)^{4}(s+9)^{2}(s+16)(s+25)\big],
\end{split}
\label{eq:m6q6}
\end{equation}
where $m_6$ is the $g^{6}$ coefficient of $M_1-1$ and $q_6$ that of
$\mathrm{Var}_{\max}/(2\pi)^{2}$.  Since $q_2=\lambda^{2}$ and $m_2=\lambda^{2}/2$, division of the
two series gives
\begin{equation}
\rho_4=\frac{m_6-\tfrac12 q_6-\rho_2q_4}{\lambda^{2}}.
\label{eq:rho4fromm6}
\end{equation}

\subsection{Effective diffusivity}  For the unwrapped coordinate of the tilted periodic dynamics
\begin{equation}\dot x=\mu[F-U'(x)]+\sqrt{2D_0}\,\xi\end{equation} with $V(x)=U(x)-Fx$ and $D_0=\mu k_BT$, the
Stratonovich effective diffusivity is \cite{Reimann01,Reimann02}
\begin{equation}
D_{\mathrm{eff}}=\frac{D_0\,L^{2}\!\int_0^{L}I_{+}(x)^{2}I_{-}(x)\,dx}
{\big(\int_0^{L}I_{+}(x)\,dx\big)^{3}},
\label{eq:Adeff}
\end{equation}
with
\begin{equation}
\begin{split}
I_{+}(x)&=e^{-\beta V(x)}\!\int_x^{x+L}e^{\beta V(y)}dy,\\
I_{-}(x)&=e^{\beta V(x)}\!\int_{x-L}^{x}e^{-\beta V(z)}dz.
\end{split}
\end{equation}
For our tilted potential
these read explicitly
\begin{equation}
\begin{split}
I_{+}(x)&=\int_0^{2\pi}e^{-ay+g[\cos(x+y)-\cos x]}dy,\\
I_{-}(x)&=\int_0^{2\pi}e^{-ay+g[\cos x-\cos(x-y)]}dy,
\end{split}
\label{eq:Ipm}
\end{equation}
and both reduce to $(1-e^{-2\pi a})/a$ at $g=0$. 

 The corresponding stationary
current is \begin{equation}J=\frac{D_0(1-e^{-\beta FL})}{\int_0^L I_{+}dx}.\end{equation}  For our potential
$\beta V(x)=g\cos x-ax$ with $L=2\pi$ one has $\beta FL=2\pi a$, and the double integral
\begin{align}
\int_0^{2\pi}I_{+}dx=&\int_0^{2\pi}dx\int_0^{2\pi}dy\,e^{-ay+g[\cos(x+y)-\cos x]}\nonumber\\
=&2\pi M_1(1-e^{-2\pi a})/a
\end{align}
 which recovers $J=aD/(2\pi M_1)$.  At $F=0$,
Eq.~(\ref{eq:Adeff}) reduces to the Lifson--Jackson result \cite{Lifson62} $D_{\mathrm{eff}}=D/I_0(g)^{2}$,
and at $g=0$ to $D$.  Expanding Eq.~(\ref{eq:Adeff}) symbolically to $O(g^{4})$ gives
\begin{equation}
\frac{D_{\mathrm{eff}}}{D}=1+c_2\,g^{2}+c_4\,g^{4}+O(g^{6}),
\label{eq:Adeffexp}
\end{equation}
with $c_2=(3a^{2}-1)/[2(1+a^{2})^{2}]$ and
$c_4=\big(\tfrac52-\tfrac{203}{8}a^{2}-\tfrac{31}{8}a^{4}+\tfrac32 a^{6}\big)
/[(1+a^{2})^{3}(a^{2}+4)^{2}]$,
whose $a\to0$ limit matches $I_0(g)^{-2}=1-g^{2}/2+5g^{4}/32+O(g^{6})$.  In the
strong-barrier regime $g\gg1$ with fixed $a$, the
Kramers escape rates across the two neighboring saddles are
\begin{equation}r_{\pm}=\mu V_0\,(1-a^{2}/(2g^{2}))\,e^{-2g\pm\pi a-a^{2}/g}/(2\pi)\end{equation} with
$\mu V_0=Dg$.  The barrier height must be evaluated at the tilted extrema
$\sin x=-a/g$, which shifts the barrier by $a^{2}/g+O(g^{-3})$ and replaces the naive
exponent $-2g\pm\pi a$ by $-2g\pm\pi a-a^{2}/g$.  The prefactor $1-a^{2}/(2g^{2})$ is the separate curvature
(attempt-frequency) correction at the same extrema.  The velocity $v=2\pi(r_{+}-r_{-})$
matches
Eq.~(\ref{eq:strongp}), and the hopping contribution to the effective diffusivity is
\begin{equation}
D_{\mathrm{eff}}=2\pi^{2}\big(r_{+}+r_{-}\big),
\label{eq:Adeffhop}
\end{equation}
 so
that, since the common prefactor of $r_\pm$ cancels, the ratio $r_+/r_-=e^{2\pi a}$ being exactly
the affinity produced per period,
\begin{equation}
M_1d_{\mathrm{eff}}=\pi a\,\frac{r_+ + r_-}{r_+ - r_-}=\pi a\coth(\pi a),
\label{eq:Adeffratio}
\end{equation}
and hence
\begin{equation}L_{\mathrm{TUR}}/\dot\sigma\to\tanh(\pi a)/(\pi a)\le1,\end{equation} the limit quoted in
Sec.~\ref{sec:tur}.  Equation~(\ref{eq:Adeffratio}) is at the same time the equality case of the
fluctuation ceiling discussed there. 

In the two-state limit the current is a renewal process with
rates $r_\pm$, which is precisely the two-reservoir model of Ref.~\cite{GuZhang25}, so that
\begin{equation}2D_J/j=(r_+ + r_-)/(r_+ - r_-)=\coth(A/2)\end{equation} with $A=2\pi a$, where $D_J=D_{\mathrm{eff}}/(2\pi)^{2}$
and $j=v/(2\pi)$ are the per-period diffusion coefficient and current.

%%%%%%%%%%%%%%%%%%%%%%%%%%%%%%%%%%%%%%%%%%%%%%%%%%%%%%%%%%%%%%

%%%%%%%%%%%%%%%%%%%%%%%%%%%%%%%%%%%%%%%%%%%%%%%%%%%%%%%%%%%%%%

\end{document}